\documentclass[orivec,runningheads]{llncs}
\usepackage[paperheight=235mm,paperwidth=155mm,textwidth=12.2cm,textheight=19.3cm,centering]{geometry}
\usepackage[hidelinks]{hyperref}
\usepackage{graphicx}
\usepackage{color,enumerate}

\DeclareSymbolFont{frenchscript}{OMS}{ztmcm}{m}{n}
\DeclareMathSymbol{\D}{\mathord}{frenchscript}{68}   
\DeclareMathSymbol{\F}{\mathord}{frenchscript}{70}   
\DeclareMathSymbol{\Pow}{\mathord}{frenchscript}{80} 
\newcommand\Act{{\cal A}}                            
\newcommand{\T}{{\cal T}}                            
\newfont{\bbb}{bbm10}                                
\newcommand{\nat}{\mbox{\bbb N}}                     
\newcommand{\reals}{\mbox{\bbb R}}                   
\newtheorem{defi}{Definition}
\newtheorem{theo}{Theorem}
\newtheorem{prop}{Proposition}
\newtheorem{exam}{Example}
\newtheorem{lemm}{Lemma}

\newenvironment{definitionRC}[2]{\begin{defi}[#1]\label{df:#2}\rm}
                           {\end{defi}}
\newenvironment{definitionR}[1]{\begin{defi}\label{df:#1}\rm}
                           {\end{defi}}
\newenvironment{theoremRC}[2]{\begin{theo}[#1]\label{thm:#2}\rm}
                           {\end{theo}}
\newenvironment{theoremR}[1]{\begin{theo}\label{thm:#1}\rm}
                           {\end{theo}}
\newenvironment{propositionRC}[2]{\begin{prop}[#1]\label{pr:#2}\rm}
                           {\end{prop}}
\newenvironment{propositionR}[1]{\begin{prop}\label{pr:#1}\rm}
                           {\end{prop}}
\newenvironment{exampleR}[1]{\begin{exam}\label{ex:#1}\rm}
                           {\end{exam}}
\newenvironment{lemmaR}[1]{\begin{lemm}\label{lem:#1}\rm}
                           {\end{lemm}}

\newcommand{\df}[1]{Def.~\ref{df:#1}}
\newcommand{\dfn}[1]{Definition~\ref{df:#1}}
\newcommand{\thm}[1]{Thm.~\ref{thm:#1}}
\newcommand{\pr}[1]{Prop.~\ref{pr:#1}}
\newcommand{\ex}[1]{Ex.~\ref{ex:#1}}
\newcommand{\lem}[1]{Lemma~\ref{lem:#1}}
\newcommand{\sect}[1]{Section~\ref{sec:#1}}
\newcommand{\tab}[1]{Table~\ref{tab:#1}}
\newcommand{\fig}[1]{Fig.~\ref{fig:#1}}
\newcommand{\plat}[1]{\raisebox{0pt}[0pt][0pt]{#1}}  
\newcommand{\dcup}{\stackrel{\mbox{\huge .}}{\cup}}  
\def\precond#1{{\vphantom{#1}}^\bullet #1}
\def\postcond#1{{#1}^\bullet}
\newcommand{\length}{{\it length}}
\newcommand{\trace}{{\it trace}}
\renewcommand{\phi}{\varphi}
\newcommand{\obis}[2]{\mathrel{_{#1}\,		     
	\raisebox{.3ex}{$\underline{\makebox[.7em]{$\leftrightarrow$}}$}
                  \,_{#2}}}
\newcommand{\bis}[1]{\obis{}{\it #1}}		     

\begin{document}

\title{Just Testing}

\author{Rob van Glabbeek\inst{1,2}\href{https://orcid.org/0000-0003-4712-7423}{\includegraphics[scale=.04]{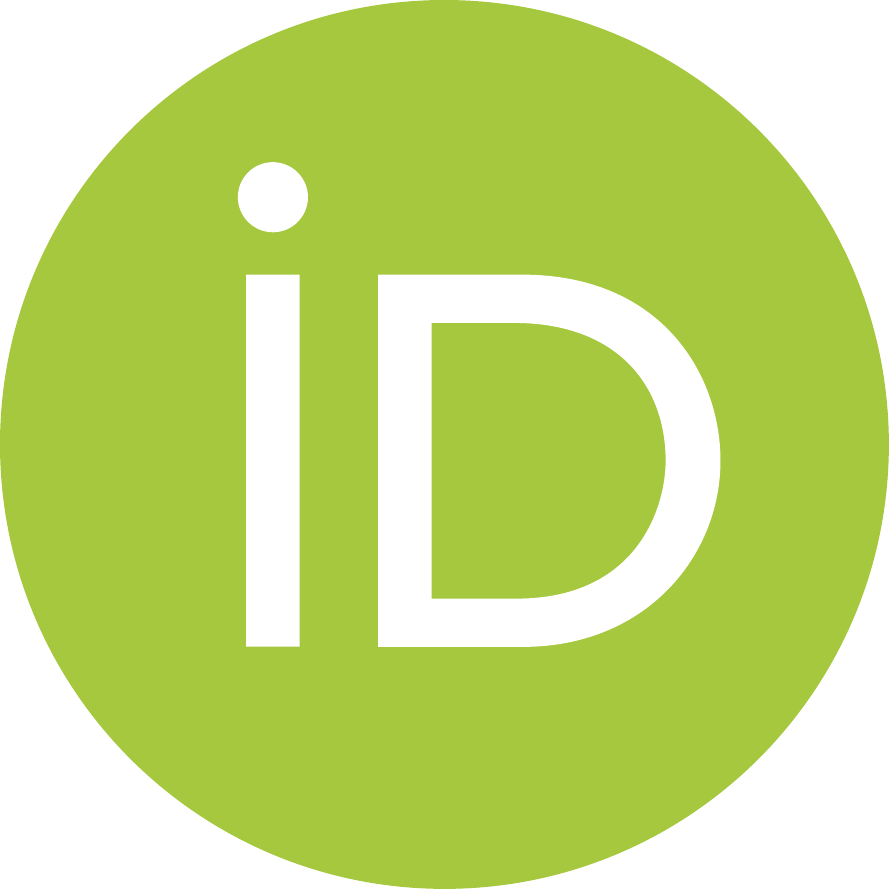}}\,%
\thanks{Supported by Royal Society Wolfson Fellowship RSWF{\textbackslash}R1{\textbackslash}221008}}
\authorrunning{R.J. van Glabbeek}
\institute{School of Informatics, University of Edinburgh\\
\and School of Comp.\ Sc.\ and Engineering, University of New South Wales, Sydney\\
\email{rvg@cs.stanford.edu}}

\maketitle
\setcounter{footnote}{0}

\begin{abstract}
The concept of must testing is naturally parametrised with a chosen completeness criterion, defining
the complete runs of a system.  Here I employ justness as this completeness criterion, instead of
the traditional choice of progress. The resulting must-testing preorder is incomparable with
the default one, and can be characterised as the fair failure preorder of Vogler.
It also is the coarsest precongruence preserving linear time properties when assuming justness.

\hspace{1em}
As my system model I here employ Petri nets with read arcs. Through their Petri net semantics, this
work applies equally well to process algebras. I provide a Petri net semantics for a standard
process algebra extended with signals; the read arcs are necessary to capture those signals.
\end{abstract}

\section{Introduction}\label{sec:introduction}

May- and must-testing was proposed by De Nicola \& Hennessy in \cite{DH84}. It yields semantic
equivalences where two processes are distinguished if and only if they react differently on certain
tests. The tests are processes that additionally feature success states. A test $\T$ is applied
to a process $N$ by taking the CCS parallel composition $\T|N$, and implicitly applying a CCS
restriction operator to it that removes the remnants of unsuccessful communication. 
Applying $\T$ to $N$ is deemed successful if and only if this composition yields a process that may,
respectively must, reach a success state.
It is trivial to recast this definition using the CSP parallel
composition $\|_\Act$ \cite{OH86} instead of the one from CCS.

It is not a priori clear how a given process \emph{must} reach a success state.
For all we know it might stay in its initial state and never take any transition leading to
this success state. To this end one must employ an assumption saying that under appropriate
circumstances certain enabled transitions will indeed be taken. Such an assumption is called a
\emph{completeness criterion} \cite{vG19}. The theory of testing from \cite{DH84} implicitly employs
a default completeness criterion that in \cite{GH19} is called \emph{progress}.
However, one can parameterise the notion of must testing by the choice of any completeness
criterion, such as the many notions of \emph{fairness} classified in \cite{GH19}.
Here I employ \emph{justness}, a completeness criterion that is better justified than
either progress or fairness \cite{GH19}.

The resulting must-testing equivalence is incomparable to the progress-based one from \cite{DH84}.
On the one hand, it no longer distinguishes deadlock and livelock, i.e., the Petri nets $N$ and $N'$
of \ex{no congruence}; on the other hand, it keeps recording information past a divergence.
I characterise the corresponding preorder as the fair failure preorder of Vogler \cite{Vog02},
which using my terminology ought to be called the \emph{just failures preorder}.
I show that it also is the coarsest precongruence preserving linear time properties when assuming justness.
Finally I show that the same preorder originates from the timed must-testing framework explored in
\cite{Vog02}, but only if all quantitative information is removed from that approach.

I carry out this work within the model of Petri nets extended with read arcs \cite{MR95,BP96}, so
that it also applies to process algebras through their standard Petri net semantics. The extension
with read arcs is necessary to capture \emph{signalling}, a process algebra operator that cannot be
adequately modelled by standard Petri nets. Signalling, or read arcs, can be used to accurately
model mutual exclusion without making a fairness assumption \cite{Vog02,CDV09,EPTCS255.2}. This is
not possible in standard Petri nets \cite{KW97,Vog02,GH15b}, or in process algebras with a standard
Petri net semantics \cite{GH15b}. Here I give a Petri net semantics of signalling, and illustrate
its use in modelling a traffic light, interacting with passing cars.

\subsubsection{Acknowledgement} I am grateful to Weiyou Wang for valuable feedback.

\section{Labelled Petri nets with read arcs}

I will employ the following notations for multisets.

\begin{definitionR}{multiset}
  Let $X$ be a set.\vspace{-1.5ex}
\begin{itemize}
\item A {\em multiset} over $X$ is a function $A\!:X \rightarrow \nat$,
i.e.\ $A\in \nat^{X}\!$.
\item $x \in X$ is an \emph{element of} $A$, notation $x \in A$, iff $A(x) > 0$.
\item For multisets $A$ and $B$ over $X$ I write $A \subseteq B$ iff
 \mbox{$A(x) \leq B(x)$} for all $x \mathbin\in X$;
\\ $A\cup B$ denotes the multiset over $X$ with $(A\cup B)(x):=\textrm{max}(A(x), B(x))$,
\\ $A\cap B$ denotes the multiset over $X$ with $(A\cap B)(x):=\textrm{min}(A(x), B(x))$,
\\ $A + B$ denotes the multiset over $X$ with $(A + B)(x):=A(x)+B(x)$,
\\ $A - B$ is given by
$(A - B)(x):=\mbox{max}(A(x)-B(x),0)$, and\\
for $k\mathbin\in\nat$ the multiset $k\cdot A$ is given by
$(k \cdot A)(x):=k\cdot A(x)$.
\item The function $\emptyset\!:X\rightarrow\nat$, given by
  $\emptyset(x):=0$ for all $x \mathbin\in X$, is the \emph{empty} multiset over $X$.
\item The cardinality $|A|$ of a multiset $A$ over $X$ is given by
  $|A| := \sum_{x\in X}A(x)$.
\item A multiset $A$ over $X$ is \emph{finite}
  iff $|A|<\infty$, i.e.,
  iff the set $\{x \mid x \mathbin\in A\}$ is finite.
\end{itemize}
With $\{x,x,y\}$ I denote the multiset over $\{x,y\}$ with
$A(x)\mathbin=2$ and $A(y)\mathbin=1$, rather than the set $\{x,y\}$ itself.
A multiset $A$ with $A(x) \leq 1$ for all $x$ is
identified with the set $\{x \mid A(x)=1\}$.
\end{definitionR}
I employ general labelled place/transition systems extended with read arcs \cite{MR95,BP96}.

\begin{definitionR}{Petri net}
  Let $\Act$ be a set of \emph{visible actions} and
  $\tau\mathbin{\not\in}\Act$ be an \emph{invisible action}.
  Let $\Act_\tau\mathbin{:=}\Act \dcup \{\tau\}$.
  A (\emph{labelled}) \emph{Petri net} (\emph{over $\Act_\tau$}) is a tuple
  $(S, T, F, R, M_0, \ell)$ where
  \begin{list}{{\bf --}}{\leftmargin 18pt
                        \labelwidth\leftmargini\advance\labelwidth-\labelsep
                        \topsep 0pt \itemsep 0pt \parsep 0pt}
    \item $S$ and $T$ are disjoint sets (of \emph{places} and \emph{transitions}),
    \item $F: ((S \times T) \cup (T \times S)) \rightarrow \nat$
      (the \emph{flow relation} including \emph{arc weights}),
    \item $R: S \times T \rightarrow \nat$ (the \emph{read} relation),
    \item $M_0 : S \rightarrow \nat$ (the \emph{initial marking}), and
    \item \plat{$\ell: T \rightarrow \Act_\tau$} (the \emph{labelling function}).
  \end{list}
\end{definitionR}

\noindent
Petri nets are depicted by drawing the places as circles and the
transitions as boxes, containing their label.  Identities of places
and transitions are displayed next to the net element.  When
$F(x,y)>0$ for $x,y \mathbin\in S\cup T$ there is an arrow (\emph{arc})
from $x$ to $y$, labelled with the \emph{arc weight} $F(x,y)$.
Weights 1 are elided.  An element $(s,t)$ of the multiset $R$ is
called a \emph{read arc}. Read arcs are drawn as lines without arrowhead.
When a Petri net represents a concurrent system, a global state of
this system is given as a \emph{marking}, a multiset $M$ of places,
depicted by placing $M(s)$ dots (\emph{tokens}) in each place $s$.
The initial state is $M_0$.

The behaviour of a Petri net is defined by the possible moves between
markings $M$ and $M'$, which take place when a finite multiset $G$ of
transitions \emph{fires}.  In that case, each occurrence of a
transition $t$ in $G$ consumes $F(s,t)$ tokens from each 
place $s$.  Naturally, this can happen only if $M$ makes all these
tokens available in the first place. Moreover, for each $t\in G$ there
need to be at least $R(s,t)$ tokens in each place $s$ that
are not consumed when firing $G$. Next, each $t$ produces $F(t,s)$
tokens in each place $s$.  \dfn{firing} formalises this notion of behaviour.

\begin{definitionR}{preset}
  Let $N = (S, T, F, R, M_0, \ell)$ be a Petri net.
    The multisets $\widehat{t},~\precond{t},~\postcond{t}: S \rightarrow\nat$ are given by
    $\widehat{t}(s)=R(s,t)$, $\precond{t}(s)=F(s,t)$ and
$\postcond{t}(s)=F(t,s)$ for all $s \mathbin\in S$.
The elements of $\widehat{t}$, $\precond{t}$ and $\postcond{t}$ are
called \emph{read-}, \emph{pre-} and \emph{postplaces} of $t$, respectively.
These functions extend to finite multisets
$G{:}\; T \rightarrow\nat$ by
$\widehat{G} \mathbin{:=} \bigcup_{t \in G}\;\widehat{t}$,~
$\precond{\!G} \mathbin{:=} \sum_{t \in T}G(t)\cdot\precond{t}$
and $\postcond{G} \mathbin{:=} \sum_{t \in T}G(t)\cdot\postcond{t}\!$.
\end{definitionR}

\begin{definitionRC}{\cite{BP96}}{firing}
Let $N \mathbin= (S, T, F, R, M_0, \ell)$ be a Petri net,
$G \mathbin\in \nat^T\!$ non-empty and finite, and $M, M' \in \nat^S$.
  $G$ is a \emph{step} from $M$ to $M'$,
written \plat{$M~[G\rangle_N~ M'$}, iff
\begin{list}{{\bf --}}{\leftmargin 18pt
                        \labelwidth\leftmargini\advance\labelwidth-\labelsep
                        \topsep 0pt \itemsep 0pt \parsep 0pt}
  \item $\precond{G} + \widehat{G} \subseteq M$ ($G$ is \emph{enabled}) and
  \item $M' = (M - \precond{G}) + \postcond{G}$.
\end{list}
\end{definitionRC}
Note that steps are (finite) multisets, thus allowing self-concurrency,
i.e.\ the same transition can occur multiple times in a single step.
One writes $M~[t\rangle_N~ M'$ for $M\mathrel{[\{t\}\rangle_N} M'$, whereas
$M [t\rangle_N$ abbreviates $\exists M'.~ M \mathrel{[t\rangle_N} M'$.
The subscript $N$ may be omitted if clear from context.

In my Petri nets transitions are labelled with \emph{actions} drawn from a
set \plat{$\Act \dcup \{\tau\}$}. This makes it possible to see these
nets as models of \emph{reactive systems} that interact with their
environment. A transition $t$ can be thought of as the occurrence of
the action $\ell(t)$. If $\ell(t)\mathbin\in\Act$, this occurrence can be
observed and influenced by the environment, but if $\ell(t)\mathbin=\tau$,
it cannot and $t$ is an \emph{internal} or \emph{silent} transition.\pagebreak[3]
Transitions whose occurrences cannot be distinguished by the
environment carry the same label. In particular, since
the environment cannot observe the occurrence of internal
transitions at all, they are all labelled $\tau$.

In \cite{KW97,Vog02,GH15b} it was established that mutual exclusion protocols cannot be correctly
modelled in standard Petri nets (without read arcs, i.e., satisfying $R(s,t)\mathbin=0$ for all
$s\mathbin\in S$ and $t\mathbin\in T$), unless their correctness becomes contingent on
making a fairness assumption. In \cite{GH15b} it was concluded from this that mutual exclusion
protocols can likewise not be correctly expressed in standard process algebras such as 
CCS \cite{Mi89}, CSP \cite{BHR84} or ACP \cite{BK85}, at least when sticking to their standard
Petri net semantics. Yet Vogler showed that mutual exclusion can be correctly modelled in Petri nets
with read arcs \cite{Vog02}, and \cite{CDV09,EPTCS255.2} demonstrate how mutual exclusion can be correctly
modelled in a process algebra extended with \emph{signalling} \cite{Be88b}.
Thus signalling adds expressiveness to process algebra that cannot be adequately
modelled in terms of standard Petri nets. This is my main reason to use Petri nets with read arcs as
system model in this paper.

In many papers on Petri nets, the sets of places and transitions are required to be finite, or at
least countable. Here I need a milder restriction, and will limit attention to nets that are
finitary in the following sense.

\begin{definitionR}{finitary}
A Petri net $N = (S, T, F, R, M_0, \ell)$ is \emph{finitary} if $M_0$ is countable,
$\postcond{t}$ is countable for all $t\in T$, and moreover the set of transitions $t$ with
$\precond{t}=\emptyset$ is countable.
\end{definitionR}

\section{A Petri net semantics of CCSP with signalling}\label{sec:ccsps}

CCSP \cite{Ol87} is a natural mix of the process algebras CCS \cite{Mi89} and CSP \cite{BHR84},
often used in connection with Petri nets. Here I will present a Petri net semantics of a version CCSPS of
CCSP enriched with \emph{signalling} \cite{Be88b}. This builds on work from~\cite{GM84,Wi84,GV87,DDM87,Ol87,Ol91};
the only novelty is the treatment of signalling.
Petri net semantics of other process algebras, like CCS \cite{Mi89}, CSP \cite{BHR84} or
ACP \cite{BK85}, are equally well known. This Petri net semantics lifts any semantic
equivalence on Petri nets to CCSPS, or to any other process algebra, so that the results of this work
apply equally well to process algebras.

CCSPS is parametrised by the choice of sets $\Act$ of visible actions
and $\cal K$ of \emph{agent identifiers}. Its syntax is given by
\[P,Q,P_i ::= \sum_{i \in I}a_i P_i ~\mbox{\Large $\,\mid\,$}~ a \triangleright \sum_{i \in I}a_i P_i
 ~\mbox{\Large $\,\mid\,$}~ P \|_A Q ~\mbox{\Large $\,\mid\,$}~ \tau_A(P) 
 ~\mbox{\Large $\,\mid\,$}~ f(P) \mbox{\Large ~$\,\mid\,$}~ K\]
with $a,a_i \mathbin\in \Act$, $A \mathbin\subseteq \Act$, $f\!:\Act \rightarrow \Act$ and $K \in \cal K$.
Here the guarded choice $\sum_{i \in I}a_i P_i$ executes one of the actions $a_i$, followed by the
process $P_i$. The process $a \triangleright P$ behaves as $P$, except that in its initial state it
it is sending the signal $a$.\footnote{The notation $a \triangleright P$ follows \cite{CDV09};
in \cite{Be88b,EPTCS255.2} this is denoted $P\mbox{\textasciicircum} a$.}~\footnotemark\linebreak[4]%
\footnotetext{Here I require $P$ to be a guarded choice in order to avoid the need for a \emph{root condition}
\cite{vG05e} to make the equivalences of this paper into congruences. This is also the reason my
language features a guarded choice, instead of action prefixing and general choice.}%
The process $P \|_A Q$ is the partially synchronous parallel
composition of processes $P$ and $Q$, where actions from $A$ can take place only when both $P$ and
$Q$ can engage in such an action, while other actions of $P$ and $Q$ occur independently. The abstraction
operator $\tau_A$ hides action from $A$ from the environment by renaming them into $\tau$, whereas
$f$ is a straightforward relabelling operator (leaving internal\vspace{1pt} actions alone). Each agent identifier
$K$ comes with a \emph{defining equation} \plat{$K\stackrel{\it def}=P$}, with $P$ a \emph{guarded}
CCSPS expression; it behaves exactly as the body of its defining equation. Here $P$ is guarded if
each occurrence of an agent identifier within $P$ lays in the scope of a guarded choice
$\sum_{i \in I}a_i P_i$ or $a \triangleright \sum_{i \in I}a_i P_i$.

A formal Petri net semantics of CCSPS, and of each of the operators $\sum$, $\triangleright$, $\|_A$,
$\tau_A$ and $f$, appears in Appendix~\ref{CCSPS}. Here I give an informal summary.

Given nets $N_i$ for $i \mathbin\in I$, the net $\sum_{i \in I}a_i N_i$ is obtained by taking
their disjoint union, but without their initial markings $(M_0)_i$, and adding a single marked place $r$,
and for each $i \in I$ a fresh transition $t_i$, labelled $a_i$, with $\precond{t_i}=\{r\}$,
$\widehat{t_i}=\emptyset$ and $\postcond(t_i)=(M_0)_i$.

The parallel composition $N\|_A N'$ is obtained out of the disjoint union of $N$ and $N'$ by
dropping from $N$ and $N'$ all transitions $t$ with $\ell(t)\in A$, and instead adding
synchronisation transitions $(t,t')$ for each pair of transitions $t$ and $t'$ from $N$ and $N'$
with $\ell(t)=\ell(t')\in A$. One has $\precond{(t,t')}:=\precond{t\rule{0pt}{8pt}}+\precond{t'}$,\vspace{1pt} and similarly for
\plat{$\widehat{(t,t')}$} and $\postcond{(t,t')}$, i.e., all arcs are inherited.

$\tau_A$ and $f$ are renaming operators that only affect the labels of transitions.

The net $a \triangleright N$ adds to the net $N$ a single transition $u$,
labelled $a$, that may fire arbitrary often, but is enabled in the initial state of $N$ only. To this
end, take $\precond{u}=\postcond{u}=\emptyset$ and $\widehat u = M_0$,
the initial marking of $N$. I apply this construction only to nets for which its initially
marked places have no incoming arcs.

\newcommand{\TF}{{\it Traffic}}
\newcommand{\TL}{{\it TL}}
\newcommand{\tr}{{\it tr}}
\newcommand{\ty}{{\it ty}}
\newcommand{\tg}{{\it tg}}
\newcommand{\drive}{{\it drive}}
\newcommand{\nil}{{\bf 0}}

\begin{figure}
\input{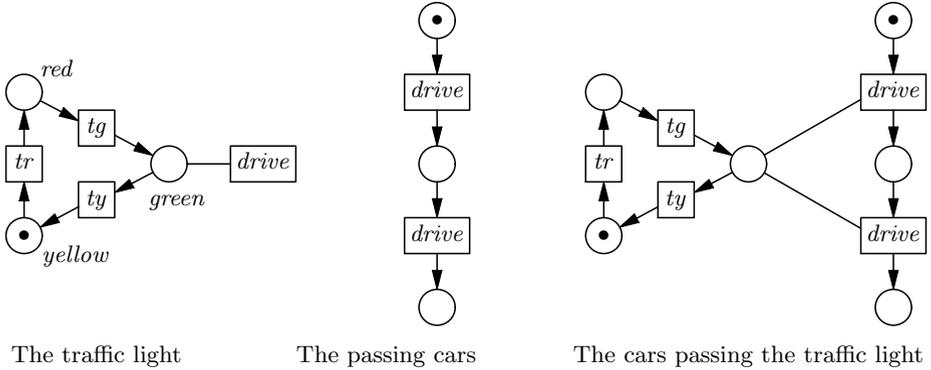}
\centerline{\box\graph}
\caption{Traffic passing traffic light}
\label{fig:traffic}\vspace{-1ex}
\end{figure}

\begin{exampleR}{traffic}
  A traffic light can be modelled by the recursive equation\vspace{-.75ex}
  $$\TL \stackrel{\it def} = \tr.\tg.(\drive \triangleright \ty.\TL).\vspace{-.75ex}$$
  Here the actions $\tr$, $\tg$ and $\ty$ stand for ``turn red'', ``turn green'' and ``turn yellow'',
  and $\drive$ indicates a state where it is OK to drive through.
  A sequence of two passing cars is modelled as
  \plat{$\TF \stackrel{\it def} = \drive.\drive.\nil$}. Here $\nil$ stands for the empty sum
  $\sum_{i\in \emptyset}a_i.E_i$ and models inaction.
  In the parallel composition $\TL ~\|_{\{\drive\}}~ \TF$ the cars only drive through when the light
  is green. All three processes are displayed in \fig{traffic}.
\vspace{-1ex}
\end{exampleR}

\section{Justness and other completeness criteria}\label{sec:cc}

\begin{definitionR}{path}
  Let $N = (S, T, F, R, M_0, \ell)$ be a Petri net.
  An \emph{execution path} $\pi$ is an alternating sequence $M_0 t_1 M_1 t_2 M_2 \dots$ of markings and
  transitions of $N$, starting with $M_0$, and either being infinite\pagebreak[3]
  or ending with a marking, such that $M_{i}~[t_{i+1}\rangle_N~ M_{i+1}$ for all $i<\length(\pi)$.
  Here $\length(\pi) \in \nat \cup\{\infty\}$ is the number of transitions in $\pi$.
  
  Let $\ell(\pi)\in \Act_\tau^\infty$ be the string $\ell(t_1) \ell(t_2) \dots$.
  Here $\Act_\tau^\infty$ denotes the collection of finite and infinite sequences of actions.
  Moreover, $\trace(\pi) \in \Act^\infty$ is obtained from $\ell(\pi)$ by dropping all occurrences of $\tau$.
 
  The execution path $\pi$ is said to \emph{enable} a transition $t$,
  notation $\pi[t\rangle$, if $M_k[ t\rangle$ for some
  $k\in\nat \wedge k \leq \length(\pi)$ and for all $k \leq j < \length(\pi)$ one has 
  $t_j \neq t$ and $(\precond{t} + \widehat t\,) \cap \precond{t_{j+1}} = \emptyset$.

  Path $\pi$ is \emph{$B$-just}, for some $B \subseteq \Act$, if $\ell(t)\in B$ for all $t\in T$ with
  $\pi[t\rangle$.
\end{definitionR}
In the definition of $\pi[ t\rangle$ above one also has $M_{j+1}[ t\rangle$ for all
$k \leq j < \length(\pi)$. Hence, a finite execution path enables a transition iff its final marking
does so. 

Informally, $\pi[t\rangle$ holds iff transition $t$ is enabled in some marking on the path $\pi$, and after
that state no transition of $\pi$ uses any of the resources needed to fire $t$. Here the read- and
preplaces of $t$ count as such resources. The clause $t_j \neq t$ moreover counts the transition
itself as one of its resources, in the sense that a transition is no longer enabled when it
occurs. This clause is redundant for transitions $t$ with $\precond{t}\neq\emptyset$. One could
interpret this clause as saying that a transition $t$ with $\precond{t}=\emptyset$ comes with
implicit marked private preplace $p_t$, and arcs $(p_t,t)$ as well as $(t,p_t)$.

\begin{figure}[b]
  \vspace{-2ex}
  \input{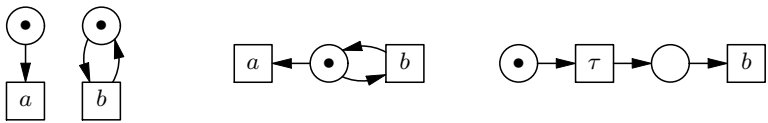}
  \centerline{\box\graph}
  \caption{(a) Progress vs.\ justness; (b) Justness vs.\ fairness; (c) $\{b\}$-progress vs.\ $\emptyset$-progress}
  \label{fig:justness}\vspace{-1ex}
\end{figure}

In \cite{vG19} I posed that Petri nets or transition systems constitute a good model of concurrency
only in combination with a \emph{completeness criterion}: a selection of a subset of all execution
paths as complete executions, modelling complete runs of the represented system.
The default completeness criterion, called \emph{progress} in \cite{GH19}, declares an execution
path complete iff it either is infinite, or its final marking enables no transition.
An alternative, called \emph{justness} in \cite{GH19}, declares an execution path complete iff it
enables no transition. Justness is a \emph{stronger} completeness
criterion than progress, in the sense that it deems fewer execution paths complete. The difference
is illustrated by the Petri net of \fig{justness}(a).
There, the execution of an infinite sequence of $b$-transitions, not involving the $a$-transition,
is complete when assuming progress, but not when assuming justness.
In the survey paper \cite{GH19}, 20 different completeness criteria are ordered by strength:
progress, justness, and 18 kinds of fairness. Most of the latter are stronger than justness:
in \fig{justness}(b) the infinite sequence of $b$-transitions is just but unfair---i.e.\ incomplete
according to these notions of fairness.
Whereas justness was a new idea in the context of transition systems \cite{GH19}, it was used as an
unnamed default assumption in much work on Petri nets \cite{Rei13}.
That justness is better warranted in applications than other completeness criteria has
been argued in \cite{GH19,vG19,GH15b,vG19c}. 

The mentioned completeness criteria from \cite{GH19} are all stronger than progress, in the sense that not all
infinite execution paths are deemed complete; on the finite execution paths they judge the same.
An orthogonal classification is obtained by varying the set $B \subseteq \Act$ of actions that may
be blocked by the environment. This fits the reactive viewpoint, in which a visible action can be
regarded as a synchronisation between the modelled system and its environment. An environment that
is not ready to synchronise with an action $b\in \Act$ can be regarded as blocking $b$.
Now \emph{$B$-progress} is the criterion that deems a path complete iff it is either infinite, or
its final marking $M$ enables only transitions with labels from $B$. When the environment may block
such transitions, it is possible for the system to not progress past $M$.
In \fig{justness}(c) the execution that performs only the $\tau$-transition is complete when
assuming $\{b\}$-progress, but not when assuming $\emptyset$-progress.
\dfn{path} defines $B$-justness accordingly, and \cite{GH19} furthermore defines 18 different notions
of $B$-fairness, for any choice of $B \subseteq \Act$. The internal action $\tau\notin B$ can
never be blocked by the environment. The default forms of progress and justness described above
correspond with $\emptyset$-progress and $\emptyset$-justness.
In \cite{Rei13} blocking and non-blocking transitions are called \emph{cold} and \emph{hot}, respectively.

\begin{figure}[h]
\input{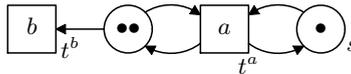}
\vspace{-2ex}
\centerline{\box\graph}\vspace{-1ex}
\caption{Run $a^\infty$ is just under the individual 
  token interpretation of Petri nets}
\label{fig:individual}
\vspace{-2ex}
\end{figure}

Two subtly different computational interpretations of Petri nets appear in the literature \cite{vG05c}:
in the \emph{individual token interpretation} multiple tokens appearing in the same place are seen as
different resources, whereas in the \emph{collective token interpretation} only the number of tokens
in a place is semantically relevant. The difference is illustrated in \fig{individual}.

\noindent
The idea underlying justness is that once a transition $t$ is enabled, eventually either $t$ will
fire, or one of the resources necessary for firing $t$ will be used by some other transition.  The
execution path $\pi$ in the net of \fig{individual} that fires the action $a$ infinitely often, but
never the action $b$, is $\emptyset$-just by \df{path}. Namely, $t^b$ is not enabled by $\pi$, as
$(\precond t^b + \widehat t^b) \cap \precond t^a \neq\emptyset$.  This fits with the individual token
interpretation, as in this run it is possible to eventually consume each token that is initially
present, and each token that stems from firing transition $t^a$.  This way any resource available for
firing $t^b$ will eventually be used by some other transition.

When adhering to the collective token interpretation of nets, execution path $\pi$ could be deemed
$\emptyset$-unjust, since transition $t^b$ can fire when there is at least one token in its preplace,
and this state of affairs can be seen as a single resource that is never taken away.
This might be formalised by adapting the definition of $\pi[t\rangle$, a path enabling a transition,
namely by changing the condition 
$(\precond{t} + \widehat t\,) \cap \precond{t_{j+1}} = \emptyset$ from \df{path}
into $\precond{t} + \widehat t + \precond{t_{j+1}} \subseteq M_j$.
However, this formalisation doesn't capture that after dropping place $s$ from the net of \fig{individual}
there is still an infinite run in which $b$ does not occur, namely when regularly firing two $a$s
simultaneously. This contradicts the conventional wisdom that firing multiple transitions at
once can always be reduced to firing them in some order. To avoid that type of complication,
I here stick to the individual token interpretation.
Alternatively, one could restrict attention to 1-safe nets \cite{Rei13}, on which there is no
difference between the individual and collective token interpretations, or to the larger class of
\emph{structural conflict nets} \cite{GGS11,vG21b}, on which the conditions
$(\precond{t} + \widehat t\,) \cap \precond{t_{j+1}} = \emptyset$ and
$\precond{t} + \widehat t + \precond{t_{j+1}} \subseteq M_j$
are equivalent \cite[Section~23.1]{vG21b}, so that \df{path} applies equally well to the collective
token interpretation.

\section{Feasibility}\label{sec:feasibility}

A standard requirement on fairness assumptions, or completeness criteria in general, is
\emph{feasibility} \cite{AFK88}, called \emph{machine closure} in \cite{Lam00}.
It says that any finite execution path can be extended into a complete one.
The following theorem shows that $B$-justness is feasible indeed.

\begin{theoremR}{feasible}
For any $B\subseteq\Act$, each finite execution path of a finitary Petri net can be extended into a $B$-just path.
\end{theoremR}
\begin{proof}
Without loss of generality I restrict attention to nets without transitions $t$ with $\precond{t}=\emptyset$.
Namely, an arbitrary net can be enriched with marked private preplaces $p_t$ for each such $t$, and arcs
$(p_t,t)$ and $(t,p_t)$. In essence, this enrichment preserves the collection of execution path of the net,
ordered by the relation ``is an extension of'', the validity of statements $\pi[t\rangle$, and the
property of $B$-justness.

  I present an algorithm extending any given path $M_0 t_1 M_1 t_2 \dots t_{k-1} M_k$ into a
  $B$-just path $\pi = M_0 t_1 M_1 t_2 M_2 \dots$.  The extension only uses transitions $t_i$
  with $\ell(t_i)\notin B$.
  As data structure my algorithm employs an $\nat\times\nat$-matrix with columns named $i$, for
  $i\geq k$, where each column has a head and a body.
  The head of column $k$ contains $M_k$ and its body lists the places $s\in M_k$,
  leaving empty most slots if there are only finitely many such places.
  Since the given net is finitary, $M_k$ has only countable many elements, so that they can be
  listed in the $\nat$ slots of column $k$.

  The head of each column $i>k$ with $i{-}1<\length(\pi)$
  will contain the pair $(t_i,M_i)$ and its body will list the places $s\in M_i$,
  again leaving empty most slots if there are only finitely many such places.
  Once more, finitariness ensures that there are enough slots in column $i$.

  An entry in the body of the matrix is either (still) empty, filled in with a place, or crossed out.
  Let $f:\nat\rightarrow \nat\times\nat$ be an enumeration of the entries in the body of this matrix.
  
  At the beginning only column $k$ is filled in; all subsequent columns of the matrix are empty.
  At each step $i> k$ I first cross out all entries $s$ in the body of the matrix
  for which there is no transition $t$ with $\ell(t)\notin B$, $M_{i-1} [t\rangle$ and $s \in \precond{t}$.
  In case all entries of the matrix are crossed out, the algorithm terminates, with output
  $M_0 t_1 M_1 t_2 \dots M_{i-1}$.
  Otherwise I fill in column $i$ as follows and cross out some more places occurring in body of the matrix.

  I take $n$ to be the smallest value such that entry $f(n)\in\nat\times\nat$ is already
  filled in, say with place $r$, but not yet crossed out.
  By the previous step of the algorithm, $M_{i-1} [t_i\rangle$ for some transition $t_i$ with
  $\ell(t_i)\notin B$ and $r \in \precond{t_i}$.
  I now fill in $(t_i,M_i)$ in the head of column $i$; here $M_i$ is the unique marking such that 
  $M_{i-1} [t_i\rangle M_i$. Subsequently I cross out all entries in the body of the matrix containing
  a place $r'\in\precond{t_i}$. This includes the entry $f(n)$.
  Finally, I fill in the body of column $i$ with the places $s\in M_i$.

  In case the algorithm doesn't terminate, the desired path $\pi$ is the sequence
  $\pi = M_0 t_1 M_1 t_2 M_2 \dots$ that is constructed in the limit.
  It remains to show that $\pi$ is $B$-just.

  Towards a contradiction, suppose $\pi[t\rangle$ for a transition $t$ with $\ell(t)\notin B$.
  By \df{path} there is an $m\in\nat \wedge m \leq \length(\pi)$ such that $M_m[t\rangle$ and
  $(\precond{t} + \widehat t\,)  \cap \precond{t_{j+1}} = \emptyset$ for all $m \leq j < \length(\pi)$.
  Let $h$ be the smallest such $m$ with $m\geq k$.
  Then there is a place $r \in \precond{t}$ appearing in column $h$.
  Here I use that $\precond{t}\neq \emptyset$.  
  This place was not yet crossed out when column $h$ was constructed.
  Since $r \notin \precond{t_{j+1}}$ and $M_{j+1}[t\rangle$ for all $h \leq j < \length(\pi)$,
  place $r$ will never be crossed out. It follows that $\pi$ must be infinite.
  The entry $r$ in column $h$ is enumerated as $f(n)$ for some $n\in \nat$, and is eventually
  reached by the algorithm and crossed out. In this regard the matrix acts as a priority queue.
  This yields the required contradiction.
\qed
\end{proof}
The above proof is a variant of \cite[Thm.~1]{vG19}, which itself is a variant of \cite[Thm.~6.1]{GH19}.
The side condition of finitariness is essential, as the below counterexample shows.

\begin{exampleR}{uncountable}
Let $N = (S, T, F, R, M_0, \ell)$ be the net with $T=\{t_r\mid r \in \reals\}$, $S=\{s_r\mid r \in \reals\}$,
$M_0(s_r)=1$, $\ell(t_r)=\tau$, $\precond{t_r}=\{s_r\}$ and $\widehat t_r = \postcond t_r= \emptyset$ for
each $r \in\reals$. It contains uncountably many action transitions, each with a marked private preplace.
As each execution path $\pi$ contains only countably many transitions, many transitions remain
enabled by $\pi$.
\end{exampleR}

\section{\hspace{-2pt}The coarsest preorders preserving linear time properties}\label{sec:coarsest}

A \emph{linear time property} is a predicate on system runs, and thus also on the execution
paths of Petri nets. One writes $\pi \models \phi$ if the execution path $\pi$ satisfies the 
linear-time property $\phi$. As the observable behaviour of an execution path $\pi$ of a Petri net
is deemed to be $\trace(\pi)$, in this context one studies only linear time properties $\phi$ such that
\begin{equation}\label{observable content}
  \trace(\pi)=\trace(\pi') ~~\Leftrightarrow~~ (\pi \models \phi \Leftrightarrow \pi' \models \phi)\;.
\end{equation}
For this reason, a linear time property can be defined or characterised as a subset of $\Act^\infty$.

Linear time properties can be used to formalise correctness requirements on systems.
They are deemed to hold for (or be satisfied by) a system iff they hold for all its complete runs.
Following \cite{EPTCS322.6} I write $\D \models^{CC} \phi$ iff property $\phi$
holds for all runs of the distributed system $\D$---and $N \models^{CC} \phi$ iff it holds for 
all execution paths of the Petri net $N$---that are complete according to the completeness criterion $CC$.
Prior to \cite{EPTCS322.6}, $\models$ was a binary predicate predicate between systems---or system
representations such as Petri nets---and properties; in this setting the default completeness
criterion of \sect{cc} was used. When using a completeness criterion {\it B-C}, where $C$ is one of the
20 completeness criteria classified in \cite{GH19} and $B \subseteq \Act$ is a modifier of $C$ based
on the set $B$ of actions that may be blocked by the environment, $N \models^{\mbox{\scriptsize\it B-C}} \phi$
is written $N \models_{B}^C \phi$ \cite{EPTCS322.6}. In this paper I am mostly interested in the values
{\it Pr} and $J$ of $C$, standing for progress and justness, respectively. To be consistent with
previous work on temporal logic, $N \models \phi$ is a shorthand for $N \models_\emptyset^{\it Pr} \phi$.

For each completeness criterion {\it B-C}, let $\sqsubseteq^C_B$ be the coarsest preorder that
preserves linear time properties when assuming {\it B-C}. Moreover, $\sqsubseteq^C$ is the coarsest
preorder that preserves linear time properties when assuming completeness criterion {\it C} in each
environment, meaning regardless which set of actions $B$ can be blocked.

\begin{definitionR}{ltpreo}
Write $N \sqsubseteq^C_B N'$ iff $N \models_{B}^C \phi ~\Rightarrow~ N' \models_{B}^C \phi$ for all linear
time properties $\phi$.
Write $N \sqsubseteq^C N'$ iff $N \sqsubseteq^C_B N'$ for all $B \subseteq \Act$.
\end{definitionR}
It is trivial to give a more explicit characterisation of these preorders.
To preserve the analogy with the failure pairs of CSP \cite{BHR84}, instead of sets $B \subseteq \Act$
I will record their complements $\overline B := \Act{\setminus} B$. As \plat{$\overline{\overline{B}}=B$},
such sets carry the same information. Since $B$ contains the actions that \emph{may} be blocked by
the environment, meaning that we consider environments that in any state may decide which actions
from $B$ to block, the set $\overline B\cup \{\tau\}$ contains actions that may not be blocked by the environment.
This means that we only consider environments that in any state are willing to synchronise with
any action in $\overline B$.

\begin{definitionR}{failures}
For completeness criterion $C$, $B$ ranging over $\Pow(\Act)$,
and Petri net $N$, let\vspace{-1ex}
\[\begin{array}{c@{\,:=\,}c@{\,|\,}l@{}}
  \F^C(N) & \{(\sigma,\overline B) & \mbox{$N$ has a $B$-$C$-complete execution path $\pi$ with
  $\sigma\mathord=\trace(\pi)$}\} \\
  \F^C_B(N) & \{~~\sigma & \mbox{$N$ has a $B$-$C$-complete execution path $\pi$ with
  $\sigma\mathord=\trace(\pi)$}\}.
\end{array}\]
\end{definitionR}
An element $(\sigma,X)$ of $\F^C(N)$ could be called a \emph{$C$-failure pair} of $N$,
because it indicates that the system represented by $N$, when executing a path with visible content
$\sigma$, may fail to execute additional actions from $X$, even when all these actions are offered by
the environment, in the sense that the environment is perpetually willing to partake in those actions.
Note that if $(\sigma,X)\in\F^C(N)$ and $Y \subseteq X$ then $(\sigma,Y)\in\F^C(N)$.

\begin{propositionR}{failures}
$N \mathbin\sqsubseteq^C_B N'$ iff $\F^C_B(N) \mathbin\supseteq \F^C_B(N')$.
\\
Likewise, $N \mathbin\sqsubseteq^C N'$ iff $\F^C(N) \mathbin\supseteq \F^C(N')$.
\end{propositionR}

\begin{proof}
  Suppose $N \sqsubseteq^C_B N'$ and $\sigma \notin \F^C_B(N)$.
  Let $\phi$ be the linear time property satisfying $\pi \models \phi$ iff $\trace(\pi) \neq \sigma$.
  Then $N \models^C_B \phi$ and thus $N' \models^C_B \phi$. Hence $\sigma \notin \F^C_B(N')$.

  Suppose $N \not\sqsubseteq^C_B N'$. There there exists a linear time property $\phi$ such that
  $N \models^C_B \phi$, yet $N' \not\models^C_B \phi$. Let $\pi'$ be a $B$-$C$-complete execution
  path of $N'$ such that $\pi'\not\models\phi$, and let $\sigma=\trace(\pi')$.
  By (\ref{observable content}) $\pi\not\models\phi$ for any execution path $\pi$ (of any net)
  such that $\trace(\pi)=\sigma$. Hence  $\sigma \in \F^C_B(N')$, yet
  $\sigma \notin \F^C_B(N)$. It follows that $\F^C_B(N) \not\supseteq \F^C_B(N')$.

  The second statement follows as a corollary of the first, using that
  $\F^C(N) \supseteq \F^C(N')$ iff
  $\F^C_B(N) \supseteq \F^C_B(N')$ for all $B\subseteq \Act$.
\qed
\end{proof}
The preorders $\sqsubseteq^C_B$ can be classified as linear time semantics \cite{vG93},
as they are characterised through reverse trace inclusions. The preorders $\sqsubseteq^C$
on the other hand capture a minimal degree of branching time. This is because they should be ready
for different choices of a system's environment at runtime.

Note that $\sqsubseteq^C$ is contained in $\sqsubseteq^C_B$ for each $B\subseteq\Act$, in the sense
that $N \sqsubseteq^C N'$ implies $N \sqsubseteq^C_B N'$. There is a priori no reason to assume
inclusions between preorders  $\sqsubseteq^C$ and $\sqsubseteq^D$ when $D$ is a stronger completeness
criterion than $C$.

To relate the preorders $\sqsubseteq^C_B$ and $\sqsubseteq^C$ with ones established in the
literature, I consider the case $C \mathbin= {\it Pr}$, i.e., taking progress as the completeness criterion~$C$.
The preorder $\sqsubseteq^{\it Pr}_\emptyset$ is
characterised as reverse inclusion of complete traces, where completeness is w.r.t.\ the default
completeness criterion of \sect{cc}. These complete traces include\vspace{-1ex}
\begin{itemize}
\item the infinite traces of a system,
\item its \emph{divergence traces} (stemming from execution paths that end in infinitely many
  $\tau$-transitions), and 
\item its \emph{deadlock traces} (stemming from finite execution paths that end in a marking enabling no
  transitions).
\vspace{-1ex}
\end{itemize}
Deadlock and divergence traces are not distinguished.
This corresponds with what is called \emph{divergence sensitive trace semantics} ($T^\lambda$) in \cite{vG93}.
The above concept of complete traces of a process $p$ is the same as in \cite{vG10}, there
denoted~${\it CT}(p)$.

The preorder $\sqsubseteq^{\it Pr}_\Act$ is characterised as reverse inclusion of infinite and
partial traces, i.e., the traces of \emph{all} execution paths.
This corresponds with what is called \emph{infinitary trace semantics} ($T^\infty$) in \cite{vG93}.
It is strictly coarser (making more identifications) than $T^\lambda$.

To analyse the preorder $\sqsubseteq^{\it Pr}$, one has $(\sigma,X)\in\F^{\it Pr}(N)$ if either\vspace{-1ex}
\begin{itemize}
\item $\sigma$ is an infinite trace of $N$---the set $X$ plays no r\^ole in that case,
\item $\sigma$ is a divergence trace of $N$, or
\item $\sigma$ is the trace of a finite path of $N$ whose end-marking enables no transition $t$
  with $\ell(t)\in X$.
\vspace{-1ex}
\end{itemize}
The resulting preorder does not occur in \cite{vG93}---it can be placed strictly between 
\emph{divergence sensitive failure semantics} ($F^\Delta$) and
\emph{divergence sensitive trace semantics} ($T^\lambda$).

The entire family of preorders $\sqsubseteq^C_B$ and $\sqsubseteq^C$ proposed in this section
was inspired by its most interesting family member, $\sqsubseteq^J$ (i.e., taking justness as the
completeness criterion $C$), proposed earlier by Walter Vogler
\cite[Def.~5.6]{Vog02}, also on Petri nets with read arcs.
Vogler \cite{Vog02} uses the word {\it fair} for what I call {\it just}.
I believe the choice of the word ``just'' is warranted to distinguish the concept from the many other
kinds of fairness that appear in the literature, which are all of a very different nature.
Accordingly, Vogler calls the semantics induced by $\sqsubseteq^J$ the {\it fair failure} semantics,
whereas I call it the \emph{just failures} semantics. My set $\F^J(N)$ is called $\F\F(N)$ in \cite{Vog02},
and Vogler addresses $\sqsupseteq^J$ simply as $\F\F$-inclusion, thereby defining it via the
right-hand side of \pr{failures}.

\section{Congruence properties}\label{sec:congruence}

A preorder $\sqsubseteq$ is called a \emph{precongruence} for an $n$-ary operator ${\it Op}$,
if $N_i \sqsubseteq N'_i$ for $i=1,\dots,n$ implies that
${\it Op}(N_1,\dots,N_n) \sqsubseteq {\it Op}(N'_1,\dots,N'_n)$. In this case the operator ${\it Op}$
is said to be \emph{monotone} w.r.t.\ the preorder $\sqsubseteq$. Being a precongruence for
important operators is known to be a valuable tool in compositional verification \cite{RBHHLPZ01}.

I write $\equiv$ for the kernel of $\sqsubseteq$, that is,
$N \equiv N'$ iff $N\sqsubseteq N' \wedge N' \sqsubseteq N$.
Here I also imply that $\equiv_B^C$ is the kernel of  $\sqsubseteq^C_B$.
If $\sqsubseteq$ is a precongruence for ${\it Op}$, then 
$\equiv$ is a \emph{congruence} for ${\it Op}$, meaning that
$N_i \equiv N'_i$ for $i=1,\dots,n$ implies that
${\it Op}(N_1,\dots,N_n) \equiv {\it Op}(N'_1,\dots,N'_n)$.

The preorder $\sqsubseteq^{\it Pr}_\Act$, characterised as reverse
inclusion of infinite and partial traces, is well-known to be precongruence for the operators of CCSP\@.
However, none of the other preorders $\sqsubseteq^{\it Pr}_B$, nor $\sqsubseteq^{\it Pr}$, is a
precongruence for parallel composition.

\begin{exampleR}{no congruence}
  Let $N=$
  \expandafter\ifx\csname graph\endcsname\relax
   \csname newbox\expandafter\endcsname\csname graph\endcsname
\fi
\ifx\graphtemp\undefined
  \csname newdimen\endcsname\graphtemp
\fi
\expandafter\setbox\csname graph\endcsname
 =\vtop{\vskip 0pt\hbox{%
\pdfliteral{
q [] 0 d 1 J 1 j
0.576 w
0.576 w
13.536 -6.768 m
13.536 -10.505863 10.505863 -13.536 6.768 -13.536 c
3.030137 -13.536 0 -10.505863 0 -6.768 c
0 -3.030137 3.030137 0 6.768 0 c
10.505863 0 13.536 -3.030137 13.536 -6.768 c
S
Q
}%
    \graphtemp=.5ex
    \advance\graphtemp by 0.094in
    \rlap{\kern 0.094in\lower\graphtemp\hbox to 0pt{\hss $\bullet$\hss}}%
    \hbox{\vrule depth0.188in width0pt height 0pt}%
    \kern 0.188in
  }%
}%
 
  \raisebox{1em}{\box\graph}
  , $N'=$
  \expandafter\ifx\csname graph\endcsname\relax
   \csname newbox\expandafter\endcsname\csname graph\endcsname
\fi
\ifx\graphtemp\undefined
  \csname newdimen\endcsname\graphtemp
\fi
\expandafter\setbox\csname graph\endcsname
 =\vtop{\vskip 0pt\hbox{%
\pdfliteral{
q [] 0 d 1 J 1 j
0.576 w
0.576 w
13.536 -6.768 m
13.536 -10.505863 10.505863 -13.536 6.768 -13.536 c
3.030137 -13.536 0 -10.505863 0 -6.768 c
0 -3.030137 3.030137 0 6.768 0 c
10.505863 0 13.536 -3.030137 13.536 -6.768 c
S
Q
}%
    \graphtemp=.5ex
    \advance\graphtemp by 0.094in
    \rlap{\kern 0.094in\lower\graphtemp\hbox to 0pt{\hss $\bullet$\hss}}%
\pdfliteral{
q [] 0 d 1 J 1 j
0.576 w
27.072 -13.536 m
40.608 -13.536 l
40.608 0 l
27.072 0 l
27.072 -13.536 l
S
Q
}%
    \graphtemp=.5ex
    \advance\graphtemp by 0.094in
    \rlap{\kern 0.470in\lower\graphtemp\hbox to 0pt{\hss $\tau$\hss}}%
\pdfliteral{
q [] 0 d 1 J 1 j
0.576 w
0.072 w
q 0 g
20.664 1.008 m
27.072 -2.736 l
19.656 -2.376 l
20.664 1.008 l
B Q
0.576 w
11.575794 -1.992963 m
14.310512 -0.838555 17.288282 -0.378387 20.243902 -0.653441 c
S
0.072 w
q 0 g
18.288 -14.76 m
11.52 -11.52 l
18.936 -11.232 l
18.288 -14.76 l
B Q
0.576 w
27.091612 -10.890258 m
24.476417 -12.294573 21.554512 -13.030328 18.586122 -13.031997 c
S
Q
}%
    \hbox{\vrule depth0.188in width0pt height 0pt}%
    \kern 0.564in
  }%
}%
 
  \raisebox{1em}{\box\graph}
  ~and ${\cal T}=$
  \expandafter\ifx\csname graph\endcsname\relax
   \csname newbox\expandafter\endcsname\csname graph\endcsname
\fi
\ifx\graphtemp\undefined
  \csname newdimen\endcsname\graphtemp
\fi
\expandafter\setbox\csname graph\endcsname
 =\vtop{\vskip 0pt\hbox{%
\pdfliteral{
q [] 0 d 1 J 1 j
0.576 w
0.576 w
13.536 -6.768 m
13.536 -10.505863 10.505863 -13.536 6.768 -13.536 c
3.030137 -13.536 0 -10.505863 0 -6.768 c
0 -3.030137 3.030137 0 6.768 0 c
10.505863 0 13.536 -3.030137 13.536 -6.768 c
S
Q
}%
    \graphtemp=.5ex
    \advance\graphtemp by 0.094in
    \rlap{\kern 0.094in\lower\graphtemp\hbox to 0pt{\hss $\bullet$\hss}}%
\pdfliteral{
q [] 0 d 1 J 1 j
0.576 w
27.072 -13.536 m
40.608 -13.536 l
40.608 0 l
27.072 0 l
27.072 -13.536 l
S
Q
}%
    \graphtemp=.5ex
    \advance\graphtemp by 0.094in
    \rlap{\kern 0.470in\lower\graphtemp\hbox to 0pt{\hss $w$\hss}}%
\pdfliteral{
q [] 0 d 1 J 1 j
0.576 w
0.072 w
q 0 g
19.872 -4.968 m
27.072 -6.768 l
19.872 -8.568 l
19.872 -4.968 l
B Q
0.576 w
13.536 -6.768 m
19.872 -6.768 l
S
Q
}%
    \hbox{\vrule depth0.188in width0pt height 0pt}%
    \kern 0.564in
  }%
}%
 
  \raisebox{1em}{\box\graph}
  .\vspace{1ex} Then \df{parallel} yields
  ${\cal T}\|_\emptyset N =$
  \expandafter\ifx\csname graph\endcsname\relax
   \csname newbox\expandafter\endcsname\csname graph\endcsname
\fi
\ifx\graphtemp\undefined
  \csname newdimen\endcsname\graphtemp
\fi
\expandafter\setbox\csname graph\endcsname
 =\vtop{\vskip 0pt\hbox{%
\pdfliteral{
q [] 0 d 1 J 1 j
0.576 w
0.576 w
13.536 -6.768 m
13.536 -10.505863 10.505863 -13.536 6.768 -13.536 c
3.030137 -13.536 0 -10.505863 0 -6.768 c
0 -3.030137 3.030137 0 6.768 0 c
10.505863 0 13.536 -3.030137 13.536 -6.768 c
S
Q
}%
    \graphtemp=.5ex
    \advance\graphtemp by 0.094in
    \rlap{\kern 0.094in\lower\graphtemp\hbox to 0pt{\hss $\bullet$\hss}}%
    \hbox{\vrule depth0.188in width0pt height 0pt}%
    \kern 0.188in
  }%
}%
 
  \raisebox{1em}{\box\graph}
  \expandafter\ifx\csname graph\endcsname\relax
   \csname newbox\expandafter\endcsname\csname graph\endcsname
\fi
\ifx\graphtemp\undefined
  \csname newdimen\endcsname\graphtemp
\fi
\expandafter\setbox\csname graph\endcsname
 =\vtop{\vskip 0pt\hbox{%
\pdfliteral{
q [] 0 d 1 J 1 j
0.576 w
0.576 w
13.536 -6.768 m
13.536 -10.505863 10.505863 -13.536 6.768 -13.536 c
3.030137 -13.536 0 -10.505863 0 -6.768 c
0 -3.030137 3.030137 0 6.768 0 c
10.505863 0 13.536 -3.030137 13.536 -6.768 c
S
Q
}%
    \graphtemp=.5ex
    \advance\graphtemp by 0.094in
    \rlap{\kern 0.094in\lower\graphtemp\hbox to 0pt{\hss $\bullet$\hss}}%
\pdfliteral{
q [] 0 d 1 J 1 j
0.576 w
27.072 -13.536 m
40.608 -13.536 l
40.608 0 l
27.072 0 l
27.072 -13.536 l
S
Q
}%
    \graphtemp=.5ex
    \advance\graphtemp by 0.094in
    \rlap{\kern 0.470in\lower\graphtemp\hbox to 0pt{\hss $w$\hss}}%
\pdfliteral{
q [] 0 d 1 J 1 j
0.576 w
0.072 w
q 0 g
19.872 -4.968 m
27.072 -6.768 l
19.872 -8.568 l
19.872 -4.968 l
B Q
0.576 w
13.536 -6.768 m
19.872 -6.768 l
S
Q
}%
    \hbox{\vrule depth0.188in width0pt height 0pt}%
    \kern 0.564in
  }%
}%
 
  \raisebox{1em}{\box\graph}
  ~and ${\cal T} \|_\emptyset N' =$
  \expandafter\ifx\csname graph\endcsname\relax
   \csname newbox\expandafter\endcsname\csname graph\endcsname
\fi
\ifx\graphtemp\undefined
  \csname newdimen\endcsname\graphtemp
\fi
\expandafter\setbox\csname graph\endcsname
 =\vtop{\vskip 0pt\hbox{%
\pdfliteral{
q [] 0 d 1 J 1 j
0.576 w
0.576 w
13.536 -6.768 m
13.536 -10.505863 10.505863 -13.536 6.768 -13.536 c
3.030137 -13.536 0 -10.505863 0 -6.768 c
0 -3.030137 3.030137 0 6.768 0 c
10.505863 0 13.536 -3.030137 13.536 -6.768 c
S
Q
}%
    \graphtemp=.5ex
    \advance\graphtemp by 0.094in
    \rlap{\kern 0.094in\lower\graphtemp\hbox to 0pt{\hss $\bullet$\hss}}%
\pdfliteral{
q [] 0 d 1 J 1 j
0.576 w
27.072 -13.536 m
40.608 -13.536 l
40.608 0 l
27.072 0 l
27.072 -13.536 l
S
Q
}%
    \graphtemp=.5ex
    \advance\graphtemp by 0.094in
    \rlap{\kern 0.470in\lower\graphtemp\hbox to 0pt{\hss $\tau$\hss}}%
\pdfliteral{
q [] 0 d 1 J 1 j
0.576 w
0.072 w
q 0 g
20.664 1.008 m
27.072 -2.736 l
19.656 -2.376 l
20.664 1.008 l
B Q
0.576 w
11.575794 -1.992963 m
14.310512 -0.838555 17.288282 -0.378387 20.243902 -0.653441 c
S
0.072 w
q 0 g
18.288 -14.76 m
11.52 -11.52 l
18.936 -11.232 l
18.288 -14.76 l
B Q
0.576 w
27.091612 -10.890258 m
24.476417 -12.294573 21.554512 -13.030328 18.586122 -13.031997 c
S
Q
}%
    \hbox{\vrule depth0.188in width0pt height 0pt}%
    \kern 0.564in
  }%
}%

  \raisebox{1em}{\box\graph}
  \expandafter\ifx\csname graph\endcsname\relax
   \csname newbox\expandafter\endcsname\csname graph\endcsname
\fi
\ifx\graphtemp\undefined
  \csname newdimen\endcsname\graphtemp
\fi
\expandafter\setbox\csname graph\endcsname
 =\vtop{\vskip 0pt\hbox{%
\pdfliteral{
q [] 0 d 1 J 1 j
0.576 w
0.576 w
13.536 -6.768 m
13.536 -10.505863 10.505863 -13.536 6.768 -13.536 c
3.030137 -13.536 0 -10.505863 0 -6.768 c
0 -3.030137 3.030137 0 6.768 0 c
10.505863 0 13.536 -3.030137 13.536 -6.768 c
S
Q
}%
    \graphtemp=.5ex
    \advance\graphtemp by 0.094in
    \rlap{\kern 0.094in\lower\graphtemp\hbox to 0pt{\hss $\bullet$\hss}}%
\pdfliteral{
q [] 0 d 1 J 1 j
0.576 w
27.072 -13.536 m
40.608 -13.536 l
40.608 0 l
27.072 0 l
27.072 -13.536 l
S
Q
}%
    \graphtemp=.5ex
    \advance\graphtemp by 0.094in
    \rlap{\kern 0.470in\lower\graphtemp\hbox to 0pt{\hss $w$\hss}}%
\pdfliteral{
q [] 0 d 1 J 1 j
0.576 w
0.072 w
q 0 g
19.872 -4.968 m
27.072 -6.768 l
19.872 -8.568 l
19.872 -4.968 l
B Q
0.576 w
13.536 -6.768 m
19.872 -6.768 l
S
Q
}%
    \hbox{\vrule depth0.188in width0pt height 0pt}%
    \kern 0.564in
  }%
}%
 
  \raisebox{1em}{\box\graph}~.\vspace{1ex}
One has $N \equiv^{\it Pr} N'$, and thus also $N \equiv^{\it Pr}_B N'$, for each $B\subseteq \Act$.
Namely $\F^{\it Pr}(N) =\F^{\it Pr}(N') = \{(\varepsilon,X)\mid X \subseteq \Act\}$.
Here $\varepsilon$ denotes the empty string.
When fixing $B$ such that $B\neq\Act$ one may choose $w \notin B$.
Now $\varepsilon \in \F_B^{\it Pr}({\cal T} \|_\emptyset N')$, for this process has an infinite
execution path that avoids the $w$-transition, which generates a divergence trace $\varepsilon$.
Yet $\varepsilon \notin \F_B^{\it Pr}({\cal T} \|_\emptyset N)$.
Hence ${\cal T} \|_\emptyset N \not\sqsubseteq^{\it Pr}_B {\cal T} \|_\emptyset N'$,
and thus also ${\cal T} \|_\emptyset N \not\sqsubseteq^{\it Pr} {\cal T} \|_\emptyset N'$.
So neither $\sqsubseteq^{\it Pr}_B$ nor $\sqsubseteq^{\it Pr}$ are
precongruences for $\|_\emptyset$.
\end{exampleR}
A common solution to the problem of a preorder $\sqsubseteq$ not being a precongruence for certain operators is to
instead consider its \emph{congruence closure}, defined as the largest precongruence contained in $\sqsubseteq$.

In \cite{KV92,vG10} the congruence closure of $\sqsubseteq^{\it Pr}$
is characterised as the so-called \emph{NDFD} preorder $\sqsubseteq_{\it NDFD}$.
Here $N \sqsubseteq_{\it NDFD} N'$ iff $N \sqsubseteq^{\it Pr} N'$
(characterised in the previous section) and moreover the divergence traces of $N'$ are included in
those of $N$. As remarked in \cite{vG10}, here it does not matter whether one requires congruence
closure merely w.r.t.\ parallel composition and injective relabelling, or w.r.t.\ all operators of
CSP (or CCSP, or anything in between).

Unlike $\sqsubseteq^{\it Pr}$, the preorder $\sqsubseteq^J$ is a precongruence for parallel composition.
Although this has been proven already by Vogler~\cite{Vog02}, in Appendix~\ref{congruence} I provide
a proof that bypasses the auxiliary notion of urgent transitions, and provides more details.

\begin{propositionRC}{\cite{Vog02}}{congruence}
$\sqsubseteq^J$ is a precongruence for relabelling and abstraction.
\end{propositionRC}
\begin{proof}
  This follows since
  $\F^J(f(N)) = \{(f(\sigma),X) \mid (\sigma, f^{-1}(X))\in \F^J(N)\}$ and moreover
  $\F^J(\tau_I(N)) = \{(\tau_I(\sigma),X) \mid (\sigma, X\cup I)\in \F^J(N)\}$.  
Here $\tau_I(\sigma)$ is the result of pruning all $I$-actions from $\sigma \in\Act^\infty$.
\qed
\end{proof}
Trivially, $\sqsubseteq^J$ also is a precongruence for $\sum a_i P_i$ and $a \triangleright\sum a_i P_i$.

The preorder $\sqsubseteq^{\it J}_\Act$ can be seen to coincide with $\sqsubseteq^{\it Pr}_\Act$,
characterised as reverse inclusion of infinite and partial traces, and thus is a precongruence for
the operators of CCSP. Leaving open the case $|\Act{\setminus}B|=1$, the preorders
$\sqsubseteq^{\it J}_B$ with $|\Act{\setminus}B|\geq 2$ fail to be precongruences for 
parallel composition.

\begin{exampleR}{no congruence J}
  Take $b,c \notin B$. Let $N$, $N'$ and $\T$ be as shown in \fig{branching}.
\vspace{-3ex}
\begin{figure}
  \input{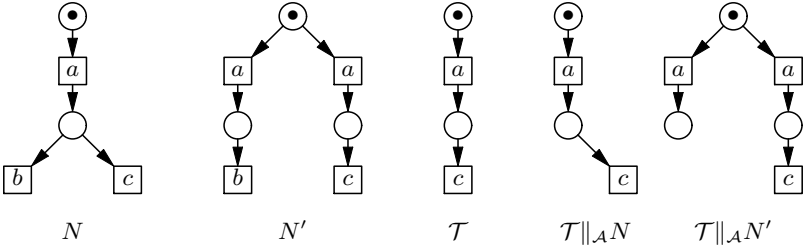}
   \centerline{\box\graph}
\caption{The preorders $\sqsubseteq^J_B$ with $|\Act{\setminus}B|\geq 2$ fail to be precongruences for parallel comp.}
\label{fig:branching}\vspace{-2.5ex}
\end{figure}
Then $N \mathbin\equiv^J_B N'$, as $\F^J_B(N)\mathbin=\F^J_B(N')\mathbin=\{\varepsilon, ab, ac\}$.
(Whether $\varepsilon$ is included depends on whether $a \mathbin\in B$.)
Yet $\T \|_\Act N \not\equiv^J_B \T\|_\Act N'$, as $a \mathbin\in \F^J_B(\T \|_\Act N')$, yet $a \mathbin{\notin} \F^J_B(\T \|_\Act N)$.
\end{exampleR}
Moreover, as illustrated below,
the preorders $\sqsubseteq^{\it J}_B$ with $B\neq \emptyset$ and $|\Act{\setminus}B|\geq 1$
fail to be precongruences for abstraction.
In the next section I will show that, for $\Act$ infinite and $B\neq\Act$, the congruence closure of
$\sqsubseteq^J_B$ for parallel composition, abstraction and relabelling is $\sqsubseteq^J$.

\begin{exampleR}{no congruence abstraction}
  Take $b \in B$ and $c\notin B$. Let $N$ and $N'$ be as shown in \fig{abstraction}.
  \begin{figure}[t]
  \input{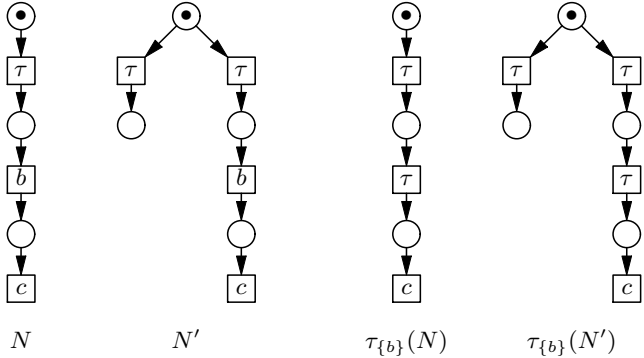}
   \centerline{\box\graph}
   \vspace{-1ex}
   \caption{The preorders $\sqsubseteq^J_B$ with $\emptyset\neq B\neq\Act$ fail to be precongruences for abstraction}
\label{fig:abstraction}\vspace{-.5ex}
\end{figure}
Then $N \equiv^J_B N'$, as $\F^J_B(N)=\F^J_B(N')=\{\varepsilon, bc\}$.
Yet $\tau_{\{b\}}(N) \not\equiv^J_B \tau_{\{b\}}(N')$, since $\varepsilon \in \F^J_B(\tau_{\{b\}}(N'))$,
yet $\varepsilon \notin \F^J_B(\tau_{\{b\}}(N))$.
\end{exampleR}
   
\section{Must Testing}

A \emph{test} is a Petri net, but featuring a special action $w\notin\Act_\tau$, not used elsewhere.
This action is used to mark \emph{success markings}: those in which $w$ is enabled.
If ${\T}$ is a test and $N$ a net then $\tau_{\Act}(\T\|_\Act N)$ is also a test.
An execution path of $\tau_{\Act}(\T\|_\Act N)$ is \emph{successful} iff it contains a success marking.
\vspace{-1pt}
\begin{definitionR}{testing}
\hspace{-2pt}A Petri net $N$ \emph{may pass} a test $\T\!$, notation $N~\textbf{may}~\T\!$, if
$\tau_{\!\Act}(\T\|_\Act\hspace{-.7pt} N)$ has a successful execution path.
It \emph{must pass} $\T$, notation $N~\textbf{must}~\T$, if each complete execution path of
$\tau_{\Act}(\T\|_\Act N)$ is successful.
It \emph{should pass} $\T$, notation $N~\textbf{should}~\T$, if each finite execution path of
$\tau_{\Act}(\T\|_\Act N)$ can be extended into a successful execution path.

Write $N \sqsubseteq_{\rm must} N'$ if
$N~\textbf{must}~\T$ implies $N'~\textbf{must}~\T$ for each test $\T$. The preorders $\sqsubseteq_{\rm may}$ and
$\sqsubseteq_{\rm should}$ are defined similarly.
\vspace{-1pt}
\end{definitionR}
The may- and must-testing preorders stem from De Nicola \& Hennessy \cite{DH84}, whereas
should-testing was added independently in \cite{BRV95} and~\cite{NC95}.

In the original work on testing \cite{DH84} the CCS parallel composition $\T|N$ was used
instead of the concealed CCSP parallel composition $\tau_{\Act}(\T\|_\Act N)$; moreover, only those
execution paths consisting solely of internal actions mattered for the definitions of passing a test.
The present approach is equivalent. First of all, restricting attention to execution paths of $\T|N$
consisting solely of internal actions is equivalent to putting $\T|N$ is the scope of a CCS
restriction operator $\backslash\Act$ \cite{Mi89}, for that operator drops all transitions of its
argument that are not labelled $\tau$ or $w$. Secondly, CCS features a complementary action $\bar a$
for each $a\in \Act$, and one has $\bar{\bar{a}}=a$. For $\T$ a test, let $\overline{\T}$ denote the
complementary test in which each action $a\in \Act$ is replaced by $\bar a$; again
\plat{$\overline{\overline{\T}}=\T$}. It follows directly from the definitions of the operators involved
that $\tau_{\Act}(\T\|_\Act N)$ is identical\footnote{The standard definition of $|$ on Petri nets
  \cite{Gol90} is given only up to isomorphism. By choosing the names of places and transitions
  similar to \df{parallel} one can obtain $\tau_{\Act}(\T\|_\Act N)=(\overline \T| N)\backslash \Act$.}
to $(\overline \T| N)\backslash \Act$.
This proves the equivalence of the two approaches.

Unlike may- and should-testing, the concept of must-testing is naturally parametrised with a
completeness criterion, deciding what counts as a complete execution. To make this choice explicit I
use the notation $\sqsubseteq_{\rm must}^C$, where $C$ could be any of the completeness
criteria surveyed in \cite{GH19}. Since processes $\tau_{\Act}(\T\|_\Act N)$ (or
$(\T| N)\backslash \Act$) do not feature any actions other than $\tau$ and $w$, where $w$ is used
merely to point to the success states, the modifier $B\subseteq\Act$ of a completeness criteria {\it B-C}
has no effect, i.e., any two choices of this modifier are equivalent.

In the original work of \cite{DH84} the default completeness criterion progress from \sect{cc} was employed.
Interestingly, $\sqsubseteq_{\rm must}^{\it Pr}$ is a congruence for the operators of CCSP that does
not preserve all linear time properties. It is strictly coarser than $\sqsubseteq_{\it NDFD}$.
In fact, it is the coarsest precongruence for the CCSP parallel composition and injective relabelling
that preserves those linear time properties that express that a system will eventually reach a state
in which something [good] has happened \cite{vG10}. (In \cite{vG10}, following \cite{Lam77}, but
deviating from the standard terminology of \cite{AS85}, such properties are called \emph{liveness properties}.)

In this paper I investigate the must-testing preorder when taking justness as the underlying completeness
criterion, $\sqsubseteq_{\rm must}^{\it J}$. \thm{just must testing} below shows that it can be
characterised as the just failures preorder $\sqsubseteq^{\it J}$ of \sect{coarsest}.

First note that \df{testing} can be simplified. When dealing with justness as completeness
criterion, the word ``complete'' in \df{testing} is instantiated by ``just'' or ``$B$-just'', for
some $B\subseteq \Act$ (not including $w$). As the result is independent of $B$, one may take
$B:=\emptyset$. Since the labelling of a net has no bearing on its execution paths, or on whether such a path
is $\emptyset$-just, or successful, one may now drop the operator $\tau_A$ from \df{testing} without
affecting the resulting notion of must testing.

\begin{theoremR}{just must testing}
  $N \sqsubseteq_{\rm must}^{\it J} N'$
  iff
  $N \sqsubseteq^J N'$.
\end{theoremR}
 
\begin{proof}
The ``if'' direction is established in Appendix~\ref{if}.

For ``only if'', suppose $N \sqsubseteq_{\rm must}^{\it J} N'$.
Using \pr{failures}, it suffices to show that $\F^J(N) \supseteq \F^J(N')$.
Let $(\sigma,X) \in \F^J(N')$, where $\sigma = a_1 a_2 \dots \in \Act^\infty$ is a finite or
infinite sequence of actions. Let $\T$ be the test displayed in \fig{universal test}.
\begin{figure}
\input{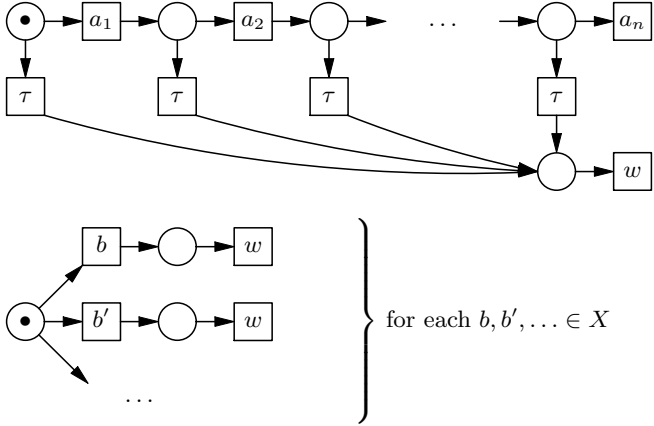}
\centerline{\box\graph}
\caption{Universal test for just must testing}
\label{fig:universal test}
\end{figure}
The drawing is for the case that  $\sigma = a_1 a_2 \dots a_n$ finite; in the infinite case, there
is no need to display $a_n$ separately.
Now $K ~\textbf{must}~ \T$, for any net $K$, when using justness as completeness criterion, iff
each $\emptyset$-just execution path of $\T\|_\Act K$ is successful, which is the case iff
$(\sigma,X)\notin \F^J(K)$.
(In other words, $\T\|_\Act K$ has an unsuccessful $\emptyset$-just execution path iff $(\sigma,X)\in \F^J(K)$.
 For the meaning of $(\sigma,X)\in \F^J(K)$ is that $K$ has
 an execution path $\pi$ with $\trace(\pi)=\sigma$ such that $\ell_K(t)\in X \Rightarrow \neg \pi[t\rangle$.)
   Hence $N' ~\textbf{must~not}~ \T$ and thus $N ~\textbf{must~not}~ \T$, and thus
   $(\sigma,X)\in\F^J(N)$.
\qed
\end{proof}

\begin{propositionR}{closure}
Let $\Act$ be infinite and $B \neq \Act$.
Then $\sqsubseteq^J$ is the congruence closure of $\sqsubseteq^J_B$ for parallel composition,
abstraction and injective relabelling.
\end{propositionR}
\begin{proof}
Pick an action $w \in \Act{\setminus}B$. Assume $N \not\sqsubseteq^J N'$. By applying an injective relabelling,
one can assure that $w$ does not occur in $N$ or $N'$. Let $(\sigma,X)\in\F^J(N')$, yet
$(\sigma,X)\notin\F^J(N)$, with $w \notin X$. Let $T$ be the net of \fig{universal test}. Then,
writing $A:=\Act{\setminus}\{w\}$,
$(\sigma,\Act)\in\F^J(\T\|_A N')$, yet $(\sigma,\Act)\notin\F^J(\T\|_A N)$.
Moreover, $(\rho,\Act)\notin \F^J(\T\|_A N')$ and $(\rho,\Act)\notin \F^J(\T\|_A N)$
for any $\rho\neq\sigma$ not containing the action $w$.
Hence, applying the proof of \pr{congruence}, using that $A \cup \overline{B} = \Act$, one has
$(\varepsilon,\overline{B})\in\F^J(\tau_{A}(\T\|_A N'))$,
yet $(\varepsilon,\overline{B})\notin\F^J(\tau_{A}(\T\|_A N))$.
Thus $\varepsilon\in\F^J_B(\tau_{A}(\T\|_A N'))$,
yet $\varepsilon\notin\F^J_B(\tau_{A}(\T\|_A N))$.
It follows that $\tau_{A}(\T\|_A N) \not\sqsubseteq^J_B \tau_{A}(\T\|_A N')$.
\qed
\end{proof}

\section{Timed must-testing}

A timed form of must-testing was proposed by Vogler in \cite{Vog02}.
Justness says that each transition that gets enabled must fire eventually, unless one of its
necessary resources will be taken away. In Vogler's framework, each transition $t$ must fire within 1
unit of time after it becomes enabled, even though it can fire faster. The implicit timer is reset
each time $t$ becomes disabled and enabled again, by another transition taken a token and returning
it to one of the replaces of $t$. Since there is no lower bound
on the time that may elapse before a transition fires, this view encompasses the same asynchronous
behaviour of nets as under the assumption of justness.

Vogler's work only pertains to \emph{safe} nets: those with the property that no reachable marking
allocates multiple tokens to the same place. Here a marking is \emph{reachable} if it occurs in some
execution path. Transitions $t$ with $\precond{t}=\emptyset$ are excluded. Although he only
considered finite nets, here I apply his work unchanged to \emph{finitely branching} nets: those in
which only finitely many transitions are enabled in each reachable marking.

\begin{definitionRC}{\cite{Vog02}}{CID}
  A \emph{continuous(ly timed ) instantaneous description (CID)} of a net $N$ is a
pair $(M,\xi)$ consisting of a marking $M$ of $N$ and a function $\xi$ mapping the transitions
enabled under $M$ to $[0, 1]$; $\xi$ describes the residual activation time of an enabled
transition.

The initial CID is CID$_0 = (M_0 ; \xi_0)$ with $\xi_0 (t) = 1$ for all $t$ with $M_0[t\rangle$.

One writes $(M,\xi)[\eta\rangle (M',\xi')$ if one of the following cases applies:\vspace{-1ex}
  \begin{enumerate}[(1)]
    \item $\eta = t \in T$, $M [t\rangle M'$, $\xi'(t):=\xi(t)$ for those transitions $t$ enabled
      under $M - \precond{t}$ and $\xi'(t):=1$ for the other transitions enabled under $M'$.
    \item $\eta = r \in \reals^+$, $r \leq \textrm{min}(\xi)$, $M' = M$ and $\xi' = \xi - r$.\vspace{-1ex}
  \end{enumerate}
  A \emph{timed execution path} $\pi$ is an alternating sequence of CIDs and elements
$t\in T$ or $r\in \reals^+$, defined just like an execution path in \df{path}.
Let $\zeta(\pi)\in \reals \cup \{\infty\}$ be the sum of all time steps in a timed execution path
$\pi$, the \emph{duration} of $\pi$.

A \emph{timed test} is a pair $(\T, D)$ of a test $\T$ and a \emph{duration} $D \in \reals_0^+$.
A net \emph{must pass} a timed test $(\T, D)$, notation $N$ \textbf{must} $(\T, D)$, if
each timed execution path $\pi$ with $\zeta(\pi)> D$ contains a transition labelled $w$.
Write $N \sqsubseteq_{\rm must}^{\rm timed} N'$ if $N~\textbf{must}~(\T,D)$ implies $N'~\textbf{must}~(\T,D)$ for each
timed test $(\T,D)$. 
\end{definitionRC}
Vogler shows that the preorder $\sqsubseteq_{\rm must}^{\rm timed}$ is strictly finer than $\sqsubseteq^J$.
In fact, although $\tau.a.\nil \equiv^J a.\nil$, one has
$\tau.a.\nil \not\equiv_{\rm must}^{\rm timed} a.\nil$, since only the latter process must pass the
timed test $(a.w, 2)$. Here I use that each of the actions
$\tau$, $a$ and $w$ may take up to 1 unit of time to occur.
A statement $N \sqsubseteq_{\rm must}^{\rm timed} N'$ says that $N'$ is
\emph{faster} than $N$, in the sense that composed with a test it is guaranteed to reach success
states in less time than $N$.

Here I show that when abstracting from the quantitative dimension of timed must-testing,
it exactly characterises $\sqsubseteq^J$.

\begin{definitionR}{must eventually}
A net \emph{must eventually pass} a test $\T$ if there exists a $D\in \reals_0^+$ such that $N$
\textbf{must} $(\T, D)$. Write $N \sqsubseteq_{\rm must}^{\rm ev.} N'$ if
when $N$ must eventually pass a test $\T$, then so does $N'$.
\end{definitionR}

\begin{theoremR}{must eventually}
Let $N,N'$ be finitely branching safe nets.
Then $N \sqsubseteq_{\rm must}^{\rm ev.} N'$ iff  $N \sqsubseteq^J N'$.
\end{theoremR}
A proof can be found in Appendix~\ref{timed}.
  
\section{Conclusion}

The just failures preorder $\sqsubseteq^J$ was introduced by Walter Vogler \cite{Vog02} in 2002.
Since then it has not received much attention in the literature, and has not been used as the
underlying semantic principle justifying actual verifications. In my view this can be seen as a
fault of the subsequent literature, as $\sqsubseteq^J$ captures exactly what is needed---no more and
no less---for the verification of safety and liveness properties of realistic systems.

I substantiate this claim by pointing out that $\sqsubseteq^J$ is the coarsest preorder preserving
safety and liveness properties when assuming justness, that is a congruence for basic process
algebra operators, such as the partially synchronous parallel composition, abstraction from internal
actions, and renaming.
As argued in \cite{GH19,vG19,GH15b,vG19c}, justness is better motivated and more suitable for
applications than competing completeness criteria, such as progress or the many notions of fairness
surveyed in \cite{GH15b}.

Moreover, I adapt the well-known must-testing preorder of De Nicola \& Hennessy \cite{DH84}, by using
justness as the underlying completeness criterion, instead of the traditional choice of progress. 
By showing that the resulting must-testing preorder $\sqsubseteq_{\rm must}^{\it J}$ coincides with 
$\sqsubseteq^J$ I strengthen the case that this is a natural and fundamental preorder.

This conclusion is further strengthened by my result that it also coincides with a qualitative
version $\sqsubseteq_{\rm must}^{\rm ev.}$ of the timed must-testing preorder $\sqsubseteq_{\rm must}^{\rm timed}$
of Vogler \cite{Vog02}. (Although $\sqsubseteq_{\rm must}^{\rm timed}$ and $\sqsubseteq^J$ stem from
the same paper~\cite{Vog02}, this connection was not made there.)

All this was shown in the setting of Petri nets extended with read arcs, and therefore also applies
to the settings of standard process algebras such as CCS, CSP or ACP\@. Since I cover read arcs, it
also applies to process algebras enriched with signalling, an operator that extends the
expressiveness of standard process algebras and is needed to accurately model mutual exclusion.
I leave it for future work to explore these matters for probabilistic models of concurrency, or
other useful extensions.

\begin{figure}
  \vspace{-8ex}
\input{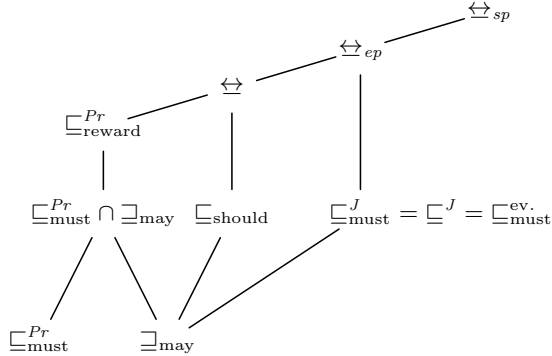}
\centerline{\box\graph}
\vspace{-1ex}
\caption{A spectrum of testing preorders and bisimilarities preserving liveness properties}
\label{fig:spectrum}
\vspace{-1.9ex}
\end{figure}

\noindent
\fig{spectrum} situates $\sqsubseteq_{\rm must}^{\it J}$ w.r.t.\ the some other semantic preorders
from the literature. The lines indicate inclusions. Here ${\sqsubseteq_{\rm must}^{\it Pr}}$,
${\sqsubseteq_{\rm may}}$ and ${\sqsubseteq_{\rm should}}$ are the classical must-, may- and
should-testing preorders from \cite{DH84} and \cite{BRV95,NC95}---see \df{testing}---and
$\sqsubseteq_{\rm reward}^{\it Pr}$ is the reward-testing preorder introduced by me in \cite{vG19b}.
The failures-divergences preorder of CSP \cite{BHR84,Ros97}, defined in a similar way
as $\sqsubseteq_{\rm must}^{\it J}$, coincides with ${\sqsubseteq_{\rm must}^{\it Pr}}$ \cite{DH84,vG19b}.
$\bis{}$ denotes the classical notion of strong bisimilarity \cite{Mi89}, and $\bis{ep}$, $\bis{sp}$
are essentially the only other preorders (in fact equivalences) that preserve linear time properties
when assuming justness: the \emph{enabling preserving bisimilarity} of \cite{GHW21ea} and the 
\emph{structure preserving bisimilarity} of \cite{vG15}.\vspace{2pt}

\noindent
\parbox{3.8in}{\hspace{10pt}The inclusions follow directly from the definitions---see refs. ---and
  counterexamples against further inclusions appear below.}\vspace{-9.4ex}

\begin{figure}
\input{counterexamples}
\hfill{\box\graph}
\end{figure}

\newpage
\bibliographystyle{splncs04}
\bibliography{refssp}

\begin{thebibliography}{10}
\providecommand{\url}[1]{\texttt{#1}}
\providecommand{\urlprefix}{URL }
\providecommand{\doi}[1]{https://doi.org/#1}

\bibitem{AS85}
Alpern, B., Schneider, F.B.: Defining liveness. Infromation Processing Letters
  \textbf{21}(4),  181--185 (1985). \doi{10.1016/0020-0190(85)90056-0}

\bibitem{AFK88}
Apt, K.R., Francez, N., Katz, S.: Appraising fairness in languages for
  distributed programming. Distributed Computing  \textbf{2}(4),  226--241
  (1988). \doi{10.1007/BF01872848}

\bibitem{Be88b}
Bergstra, J.A.: {ACP} with signals. In: Grabowski, J., Lescanne, P., Wechler,
  W. (eds.) {\rm Proc.\ International Workshop on} Algebraic and Logic
  Programming. \rm LNCS, vol.~343, pp. 11--20. Springer (1988).
  \doi{10.1007/3-540-50667-5_53}

\bibitem{BK85}
Bergstra, J.A., Klop, J.W.: Algebra of communicating processes with
  abstraction. Theor. Comput. Sci.  \textbf{37}(1),  77--121 (1985).
  \doi{10.1016/0304-3975(85)90088-X}

\bibitem{BRV95}
Brinksma, E., Rensink, A., Vogler, W.: Fair testing. In: Lee, I., Smolka, S.A.
  (eds.) {\rm Proc. 6th International Conference on} Concurrency Theory, {\rm
  {CONCUR}'95}. \rm LNCS, vol.~962, pp. 313--327. Springer (1995).
  \doi{10.1007/3-540-60218-6_23}

\bibitem{BHR84}
Brookes, S.D., Hoare, C.A.R., Roscoe, A.W.: A theory of communicating
  sequential processes. J.~ACM  \textbf{31}(3),  560--599 (1984).
  \doi{10.1145/828.833}

\bibitem{BP96}
Busi, N., Pinna, G.M.: Non sequential semantics for contextual {P/T} nets. In:
  Billington, J., Reisig, W. (eds.) {\rm Proc.\ 17th Int.\ Conf.\ on}
  Application and Theory of Petri Nets. \rm LNCS, vol.~1091, pp. 113--132.
  Springer (1996). \doi{10.1007/3-540-61363-3_7}

\bibitem{CDV09}
Corradini, F., {Di Berardini}, M.R., Vogler, W.: Time and fairness in a process
  algebra with non-blocking reading. In: Nielsen, M., Kucera, A., Miltersen,
  P.B., Palamidessi, C., Tuma, P., Valencia, F.D. (eds.) Theory and Practice of
  Computer Science, {\rm {SOFSEM}'09}. \rm LNCS, vol.~5404, pp. 193--204.
  Springer (2009). \doi{10.1007/978-3-540-95891-8_20}

\bibitem{DH84}
De~Nicola, R., Hennessy, M.: Testing equivalences for processes. Theor. Comput.
  Sci.  \textbf{34},  83--133 (1984). \doi{10.1016/0304-3975(84)90113-0}

\bibitem{DDM87}
Degano, P., {De Nicola}, R., Montanari, U.: {CCS} is an (augmented) contact
  free {C/E} system. In: {Venturini Zilli}, M. (ed.) Advanced School on
  Mathematical Models for the Semantics of Parallelism, 1986. \rm LNCS,
  vol.~280, pp. 144--165. Springer (1987). \doi{10.1007/3-540-18419-8\_13}

\bibitem{EPTCS255.2}
Dyseryn, V., van Glabbeek, R.J., H\"ofner, P.: Analysing mutual exclusion using
  process algebra with signals. In: Peters, K., Tini, S. (eds.) {\rm
  Proceedings Combined 24th International Workshop on} Expressiveness in
  Concurrency {\rm and 14th Workshop on} Structural Operational Semantics. {\rm
  EPTCS}, vol.~255, pp. 18--34 (2017). \doi{10.4204/EPTCS.255.2}

\bibitem{vG93}
van Glabbeek, R.J.: The linear time -- branching time spectrum {II}; the
  semantics of sequential systems with silent moves. In: Best, E. (ed.) {\rm
  Proc.\ CONCUR'93, 4$^{\it th}$ Int.\ Conf.\ on} Concurrency Theory. \rm LNCS,
  vol.~715, pp. 66--81. Springer (1993). \doi{10.1007/3-540-57208-2_6}

\bibitem{vG05e}
van Glabbeek, R.J.: A characterisation of weak bisimulation congruence. In:
  Middeldorp, A., van Oostrom, V., van Raamsdonk, F., de~Vrijer, R. (eds.)
  Processes, Terms and Cycles: Steps on the Road to Infinity: Essays Dedicated
  to Jan Willem Klop on the Occasion of His 60th Birthday. \rm LNCS, vol.~3838,
  pp. 26--39. Springer (2005). \doi{10.1007/11601548\_4}

\bibitem{vG05c}
van Glabbeek, R.J.: The individual and collective token interpretations of
  {Petri} nets. In: Abadi, M., de~Alfaro, L. (eds.) {\rm Proc.\ CONCUR'05,
  16$^{\it th}$ Int.\ Conf.\ on} Concurrency Theory. \rm LNCS, vol.~3653, pp.
  323--337. Springer (2005). \doi{10.1007/11539452_26}

\bibitem{vG10}
van Glabbeek, R.J.: The coarsest precongruences respecting safety and liveness
  properties. In: Calude, C., Sassone, V. (eds.) {\rm Proc.\ 6th IFIP TC 1/WG
  2.2 Int.\ Conf.\ on} Theoretical Computer Science, {\rm TCS'10; held as part
  of the {\sl World Computer Congress}}. IFIP, vol.~323, pp. 32--52. Springer
  (2010). \doi{10.1007/978-3-642-15240-5_3},
  \url{http://arxiv.org/abs/1007.5491}

\bibitem{vG15}
van Glabbeek, R.J.: Structure preserving bisimilarity, supporting an
  operational petri net semantics of {CCSP}. In: Meyer, R., Platzer, A.,
  Wehrheim, H. (eds.) {\rm Proceedings} Correct System Design - Symposium in
  Honor of Ernst-R{\"{u}}diger Olderog on the Occasion of His 60th Birthday.
  \rm LNCS, vol.~9360, pp. 99--130. Springer (2015).
  \doi{10.1007/978-3-319-23506-6_9}, \url{http://arxiv.org/abs/1509.05842}

\bibitem{vG16b}
van Glabbeek, R.J.: An algebraic treatment of recursion. In: Bethke, I.,
  Bredeweg, B., Ponse, A. (eds.) Liber Amicorum for Jan A. Bergstra, pp.
  58--59. Informatics Institute, University of Amsterdam (2016),
  \url{https://arxiv.org/abs/1702.07838}

\bibitem{vG19c}
van Glabbeek, R.J.: Ensuring liveness properties of distributed systems: Open
  problems. Journal of Logical and Algebraic Methods in Programming
  \textbf{109}, 100480 (2019). \doi{10.1016/j.jlamp.2019.100480}

\bibitem{vG19}
van Glabbeek, R.J.: Justness: A completeness criterion for capturing liveness
  properties. In: Boja\'nczyk, M., Simpson, A. (eds.) {\rm Proc.\ 22st Int.\
  Conf.\ on} Foundations of Software Science and Computation Structures, {\rm
  FoSSaCS'19; held as part of ETAPS'19}. \rm LNCS, vol. 11425, pp. 505--522.
  Springer (2019). \doi{10.1007/978-3-030-17127-8_29},
  \url{https://arxiv.org/abs/1909.00286}

\bibitem{vG19b}
van Glabbeek, R.J.: Reward testing equivalences for processes. In: Boreale, M.,
  Corradini, F., Loreti, M., Pugliese, R. (eds.) Models, Languages, and Tools
  for Concurrent and Distributed Programming, {\rm Essays Dedicated to Rocco De
  Nicola on the occasion of his 65th Birthday}, \rm LNCS, vol. 11665, pp.
  45--70. Springer (2019). \doi{10.1007/978-3-030-21485-2_5},
  \url{https://arxiv.org/abs/1907.13348}

\bibitem{EPTCS322.6}
van Glabbeek, R.J.: Reactive temporal logic. In: Dardha, O., Rot, J. (eds.)
  {\rm Proc.\ Combined 27th Int.\ Workshop on} Expressiveness in Concurrency
  {\rm and 17th Workshop on} Structural Operational Semantics. {\rm EPTCS},
  vol.~322, pp. 51--68 (2020). \doi{10.4204/EPTCS.322.6}

\bibitem{vG21b}
van Glabbeek, R.J.: Modelling mutual exclusion in a process algebra with
  time-outs (2021), \url{https://arxiv.org/abs/2106.12785}

\bibitem{GGS11}
van Glabbeek, R.J., Goltz, U., Schicke, J.W.: Abstract processes of
  place/transition systems. Information Processing Letters  \textbf{111}(13),
  626--633 (2011). \doi{10.1016/j.ipl.2011.03.013},
  \url{https://arxiv.org/abs/1103.5916}

\bibitem{GH15b}
van Glabbeek, R.J., H{\"{o}}fner, P.: {CCS:} it's not fair! -- fair schedulers
  cannot be implemented in {CCS}-like languages even under progress and certain
  fairness assumptions. Acta Informatica  \textbf{52}(2-3),  175--205 (2015).
  \doi{10.1007/s00236-015-0221-6}, \url{https://arxiv.org/abs/1505.05964}

\bibitem{GH19}
van Glabbeek, R.J., H{\"o}fner, P.: Progress, justness and fairness. ACM
  Computing Surveys  \textbf{52}(4), 69 (August 2019). \doi{10.1145/3329125},
  \url{https://arxiv.org/abs/1810.07414}

\bibitem{GHW21ea}
van Glabbeek, R.J., H{\"{o}}fner, P., Wang, W.: Enabling preserving
  bisimulation equivalence. In: Haddad, S., Varacca, D. (eds.) {\rm Proc. 32nd
  Int. Conference on} Concurrency Theory, {\rm CONCUR'21}. Leibniz
  International Proceedings in Informatics (LIPIcs), vol.~203. Schloss
  Dagstuhl--Leibniz-Zentrum f\"ur Informatik (2021).
  \doi{10.4230/LIPIcs.CONCUR.2021.33}, \url{https://arxiv.org/abs/2108.00142}

\bibitem{GV87}
van Glabbeek, R.J., Vaandrager, F.W.: Petri net models for algebraic theories
  of concurrency. In: Bakker, J.W.d., Nijman, A.J., Treleaven, P.C. (eds.) {\rm
  Proc.\ PARLE}, Parallel Architectures and Languages Europe, Vol. II. \rm
  LNCS, vol.~259, pp. 224--242. Springer (1987). \doi{10.1007/3-540-17945-3_13}

\bibitem{Gol90}
Goltz, U.: {CCS} and {Petri} nets. In: Guessarian, I. (ed.) {\rm Proc}.\
  Semantics of Systems of Concurrent Processes, {LITP} Spring School on
  Theoretical Computer Science. \rm LNCS, vol.~469, pp. 334--357. Springer
  (1990). \doi{10.1007/3-540-53479-2_14}

\bibitem{GM84}
Goltz, U., Mycroft, A.: On the relationship of {CCS} and {Petri} nets. In:
  Paredaens, J. (ed.) {\rm Proc. $11^{th}$ Colloquium on} Automata, Languages
  and Programming, {\rm ICALP84}. \rm LNCS, vol.~172, pp. 196--208. Springer
  (1984). \doi{10.1007/3-540-13345-3_18}

\bibitem{KV92}
Kaivola, R., Valmari, A.: The weakest compositional semantic equivalence
  preserving nexttime-less linear temporal logic. In: Cleaveland, R. (ed.)
  Proc.\ CONCUR'92. \rm LNCS, vol.~630, pp. 207--221. Springer (1992).
  \doi{10.1007/BFb0084793}

\bibitem{KW97}
Kindler, E., Walter, R.: Mutex needs fairness. Inf. Process. Lett.
  \textbf{62}(1),  31--39 (1997). \doi{10.1016/S0020-0190(97)00033-1}

\bibitem{Lam77}
Lamport, L.: Proving the correctness of multiprocess programs. IEEE
  Transactions on Software Engineering  \textbf{3}(2),  125--143 (1977).
  \doi{10.1109/TSE.1977.229904}

\bibitem{Lam00}
Lamport, L.: Fairness and hyperfairness. Distributed Computing  \textbf{13}(4),
   239--245 (2000). \doi{10.1007/PL00008921}

\bibitem{Mi89}
Milner, R.: Communication and Concurrency. Prentice-Hall (1989), alternatively
  see{ \em A Calculus of Communicating Systems}, LNCS 92, Springer, 1980,
  \url{https://doi.org/10.1007/3-540-10235-3}

\bibitem{MR95}
Montanari, U., Rossi, F.: Contextual nets. Acta Informatica  \textbf{32}(6),
  545--596 (1995). \doi{10.1007/BF01178907}

\bibitem{NC95}
Natarajan, V., Cleaveland, R.: Divergence and fair testing. In:
  F{\"{u}}l{\"{o}}p, Z., G{\'{e}}cseg, F. (eds.) {\rm Proc.\ 22nd Int.\
  Colloquium on} Automata, Languages and Programming, {\rm ICALP'95}. \rm LNCS,
  vol.~944, pp. 648--659. Springer (1995). \doi{10.1007/3-540-60084-1_112}

\bibitem{Ol87}
Olderog, E.R.: Operational {Petri} net semantics for {CCSP}. In: Rozenberg, G.
  (ed.) Advances in Petri Nets 1987. \rm LNCS, vol.~266, pp. 196--223. Springer
  (1987). \doi{10.1007/3-540-18086-9_27}

\bibitem{Ol91}
Olderog, E.R.: Nets, Terms and Formulas: Three Views of Concurrent Processes
  and Their Relationship. Cambridge Tracts in Theoretical Computer Science 23,
  Cambridge University Press (1991)

\bibitem{OH86}
Olderog, E.R., Hoare, C.A.R.: Specification-oriented semantics for
  communicating processes. Acta Inf.  \textbf{23},  9--66 (1986).
  \doi{10.1007/BF00268075}

\bibitem{Rei13}
Reisig, W.: Understanding Petri Nets --- Modeling Techniques, Analysis Methods,
  Case Studies. Springer (2013). \doi{10.1007/978-3-642-33278-4}

\bibitem{RBHHLPZ01}
Roever, W.P.d., de~Boer, F.S., Hannemann, U., Hooman, J., Lakhnech, Y., Poel,
  M., Zwiers, J.: Concurrency Verification: Introduction to Compositional and
  Noncompositional Methods, Cambridge Tracts in TCS, vol.~54. Cambridge
  University Press (2001)

\bibitem{Ros97}
Roscoe, A.W.: The Theory and Practice of Concurrency. Prentice-Hall (1997),
  \url{http://www.comlab.ox.ac.uk/bill.roscoe/publications/68b.pdf}

\bibitem{Vog02}
Vogler, W.: Efficiency of asynchronous systems, read arcs, and the
  {MUTEX}-problem. Theor. Comput. Sci.  \textbf{275}(1-2),  589--631 (2002).
  \doi{10.1016/S0304-3975(01)00300-0}

\bibitem{Wi84}
Winskel, G.: A new definition of morphism on {Petri} nets. In: Fontet, M.,
  Mehlhorn, K. (eds.) {\rm Proc.} Symposium of Theoretical Aspects of Computer
  Science, {\rm STACS'84}. \rm LNCS, vol.~166, pp. 140--150. Springer (1984).
  \doi{10.1007/3-540-12920-0_13}

\end{thebibliography}

\newpage
\appendix

\section{Petri net semantics of CCSPS}\label{CCSPS}

Here I interpret the operators of CCSPS in terms of Petri nets with read arcs.
The definition of $\|_A$ below is based on the one from \cite{Vog02}. When
introducing a net $N_i$, its components are understood to be $(S_i,T_i,F_i,R_i,{M_0}_i,\ell_i)$.

\begin{definitionR}{parallel}
  Let $N_1$ and $N_2$ be Petri nets and $A \subseteq \Act$.
  The \emph{parallel composition} $N = N_1 \|_A N_2$ with synchronisation over $A$ is defined by
  \begin{itemize}
  \item $S := \{(s_1,*)\mid s_1\in S_1\} \cup \{(*,s_2)\mid s_2\in S_2\}$,
  \item \(T := \begin{array}[t]{@{}l@{}}
              \{(t_1,t_2) \mid t_1\in T_1 \wedge t_2 \in T_2 \wedge \ell_1(t_1)=\ell_2(t_2)\in A\}
              \cup \mbox{}\\
              \{(t_1,*) \mid t_1 \in T_1 \wedge \ell_1(t_1)\notin A\}
              \cup
              \{(*,t_2) \mid t_2 \in T_2 \wedge \ell_2(t_2)\notin A\}
              \end{array}\)
  \item \(F((x_1,x_2),(y_1,y_2)) := \left\{\begin{array}{@{}l@{~}l@{}}
               F(x_1,y_1) &\mbox{if}~ x_1 \mathop{\neq} * \mathop{\neq} y_1 \\
               F(x_2,y_2) &\mbox{if}~ x_2 \mathop{\neq} * \mathop{\neq} y_2 \\
               0 & \mbox{otherwise}
               \end{array}\right\}\) \hfill for \hfill
    $\begin{array}{@{}l@{}}((x_1,x_2),(y_1,y_2))\\ \in S\times T\cup T \times S\end{array}$
  \item \(R((s_1,s_2),(t_1,t_2)) := \left\{\begin{array}{@{}l@{~}l@{}}
               R(s_1,t_1) &\mbox{if}~ s_1 \mathop{\neq} * \mathop{\neq} t_1 \\
               R(s_2,t_2) &\mbox{if}~ s_2 \mathop{\neq} * \mathop{\neq} t_2 \\
               0 & \mbox{otherwise}
              \end{array}\right\}\) \hfill for \hfill
    $\begin{array}{@{}l@{}}((s_1,s_2),(t_1,t_2)) \\ \in S\times T\end{array}$
  \item \(M_0((s_1,s_2)) := \left\{\begin{array}{@{}l@{~~\mbox{if}~}l@{}}
               {M_0}_1(s_1) & s_1 \in S_1 \\
               {M_0}_2(s_2) & s_2 \in S_2 \\
              \end{array}\right.\)
  \hfill and \hfill \(\ell((t_1,t_2)) := \left\{\begin{array}{@{}l@{~~\mbox{if}~}l@{}}
               \ell_1(t_1) & t_1 \in T_1 \\
               \ell_2(t_2) & t_2 \in T_2 \;.\\
              \end{array}\right.\)
  \end{itemize}
\end{definitionR}

\begin{definitionR}{relabelling}
Let $f:\Act \rightarrow \Act$ be a \emph{relabelling function};
it is extended to $\Act_\tau$ by $f(\tau)\mathbin=\tau$.
Given a net $N=(S,T,F,R,M_0,\ell)$, the net $f(N)=(S,T,F,R,M_0,f\circ\ell)$ differs only in its
labelling function. Each label $a \in \Act$ is replaced by $f(a)$.

Let $I \mathbin\subseteq \Act$. The function $\tau_I:\Act_\tau \rightarrow \Act_\tau$ is given by
$\tau_I(a)\mathbin=\tau$ if $a\mathbin\in I$ and $\tau_I(a)\mathbin=a$ otherwise.
Given $N=(S,T,F,R,M_0,\ell)$, the net $\tau_I(N)=(S,T,F,R,M_0,\tau_I\circ\ell)$ differs only in its
labelling function. Each label $a \in I$ is replaced by $\tau$.
\end{definitionR}

\begin{definitionR}{guarded choice}
Given nets $N_i$ and actions $a_i\mathbin\in\Act$ for $i\mathbin\in I\mathbin{\not\ni} *$, the Petri
net $N={\color{blue} a} \mathbin{\color{blue}\triangleright}\sum_{i\in i}a_i N_i$ is defined by
  \begin{itemize}
  \item $S := \{(s,i)\mid i \in I \wedge s \in S_i\} \cup \{(r,*)\}$
  \item $T := \{(t,i)\mid i \in I \wedge t \in T_i\} \cup \{(t_i,*)\mid i \in I\} \color{blue} \cup \{(u,\triangleright)\}$
  \item \(F((x,i),(y,j)) := \left\{\begin{array}{@{}l@{~}l@{}}
               F(x,y) &\mbox{if}~ i=j\in I \\
               1 &\mbox{if}~ x=r \wedge i=j=* \\
               {M_0}_i(y) &\mbox{if}~ (x,i)=(t_j,*) \wedge j\in I \\
               0 & \mbox{otherwise}
               \end{array}\right\}\) \hfill for \hfill
    $\begin{array}{@{}l@{}}((x,i),(y,j))\\ \in S\times T\cup T \times S\end{array}$
  \item \(R((s,i),(t,j)) := \left\{\begin{array}{@{}l@{~}l@{}}
               R(s,t) &\mbox{if}~ i=j\in I \\
               \color{blue} 1 & \color{blue}\mbox{if}~ i=* \wedge j=\triangleright\\
               0 & \mbox{otherwise}
               \end{array}\right\}\) \hfill for \hfill $((s,i),(t,j)) \in S\times T$
  \item $M((r,*)):=1$ and $M((s,i)):= 0$ for each $i \in I$ and $s \in S_i$
  \item $\ell(t,i) := \ell_i(t)$ for $i \mathbin\in I$ and $t\mathbin\in T_i$, {\color{blue} $\ell(u,\triangleright)=a$}
    and $\ell(t_i,*) := a_i$ for each $i \mathbin\in I$.
  \end{itemize}
The definition of $N=\sum_{i\in i}a_i N_i$ is the same, but skipping the blue parts.
\end{definitionR}

\newcommand{\denote}[1]{[\hspace{-1.4pt}[#1]\hspace{-1.4pt}]}  
\newcommand{\SC}{{S}}                                          
\makeatletter
\def\comesfrom{\@transition\leftarrowfill}
\def\goesto{\@transition\rightarrowfill}
\def\ngoesto{\@transition\nrightarrowfill}
\def\@transition#1{\@@transition{#1}}
\newbox\@transbox
\newbox\@arrowbox
\newbox\@downbox
\def\@@transition#1#2%
   {\setbox\@transbox\hbox
      {\vrule height 1.5ex depth .9ex width 0ex\hskip0.25em$\scriptstyle#2$\hskip0.25em}
   \ifdim\wd\@transbox<1.5em
      \setbox\@transbox\hbox to 1.5em{\hfil\box\@transbox\hfil}\fi
   \setbox\@arrowbox\hbox to \wd\@transbox{#1}
   \ht\@arrowbox\z@\dp\@arrowbox\z@
   \setbox\@transbox\hbox{$\mathop{\box\@arrowbox}\limits^{\box\@transbox}$}
   \dp\@transbox\z@\ht\@transbox 10pt
   \mathrel{\box\@transbox}}
\def\nrightarrowfill{$\m@th\mathord-\mkern-6mu%
  \cleaders\hbox{$\mkern-2mu\mathord-\mkern-2mu$}\hfill
  \mkern-6mu\mathord\not\mkern-2mu\mathord\rightarrow$}
\makeatother 
\newcommand{\ar}[1]{\mathrel{\goesto{#1}}}           
\newcommand{\nar}[1]{\mathrel{\ngoesto{#1\;}}}       

\noindent
To give a semantic interpretation of recursion I follow the operational approach of
Degano, De Nicola \& Montanari \cite{DDM87} and Olderog \cite{Ol87,Ol91}.\footnote{When aiming for a
semantics in terms of Petri nets modulo $\equiv^J$, the algebraic approach of \cite{vG16b} is an alternative.}
The standard operational semantics of process algebras like CCSPS or CCS \cite{Mi89} yields one big labelled
transition system for the entire language.\footnote{A \emph{labelled transition system} (LTS)
  is given by a set $S$ of \emph{states} and a \emph{transition relation}
  \mbox{$T\subseteq S\times \Act_\tau \times S$}.
}
Each individual CCSPS expression $P$ appears as a state in
this LTS\@.  If desired, a \emph{process graph}---an LTS enriched with an initial state---for $P$
can be extracted from this system-wide LTS by appointing $P$ as the initial state, and optionally
deleting all states and transitions not reachable from $P$. In the same vein, an operational Petri
net semantics yields one big Petri net for the entire language, but without an initial marking.
I call such a Petri net {\em unmarked}. Each CCSPS expression $P$ corresponds with a marking
$dex(P)$ of that net. If desired, a Petri net $\denote{P}$ for $P$ can be extracted from this
system-wide net by appointing $dex(P)$ as its initial marking, and optionally deleting all places
and transitions not reachable from $dex(P)$.

The set $\SC_{\rm CCSPS}$ of places in the net is the smallest set including:
\begin{center}
\begin{tabular}{@{}l@{\quad}l@{\qquad\quad}l@{\quad}l@{}}
$\sum_{i \in I}a_i P_i$ & \emph{guarded choice} &
$a \triangleright \sum_{i \in I}a_i P_i$ & \emph{signalling guarded choice} \\
$\mu\|_A$ & \emph{left component}&
$_A\|\mu$ & \emph{right parallel component} \\
$\tau_I(\mu)$ &  \emph{abstraction} &
$f(\mu)$ &  \emph{relabelling} \\
\end{tabular}
\end{center}
\noindent for $a,a_i\mathbin\in \Act$, $P_i$ CCSPS expressions, $A,I\subseteq \Act$,
$\mu,\nu\mathbin\in\SC_{\rm CCSPS}$ and relabelling functions $f$.
The mapping $dex$ from CCSPS expressions to $\Pow(\SC_{\rm CCSPS})$ decomposing and expanding a
process expression into a set of places is inductively defined by:
\[
\begin{array}{@{}l@{~=~}l@{}}
dex(\sum_{i \in I}a_i P_i) & \{\sum_{i \in I}a_i P_i\} \\
dex(a \triangleright \sum_{i \in I}a_i P_i) & \{a \triangleright \sum_{i \in I}a_i P_i\} \\
dex(P\|_A Q) & dex(P)\|_A~ \cup ~_A\|dex(Q) \\
dex(\tau_I(P)) & \tau_I(dex(P)) \\
dex(f(P)) & f(dex(P)) \\
dex(K) & dex(P) \quad \mbox{when}~ \plat{$K \stackrel{\it def} = P$}.
\end{array}
\]
Here $H\|_A$, $_A\|H$, $\tau_I(H)$ and $f(H)$ for $H,K\mathbin\subseteq \SC_{\rm CCSPS}$
are defined element by element; e.g.\ $f(H) = \{f(\mu) \mid \mu\mathbin\in H\}$.
Binding matters, so $(_A\|H)\|_B\mathbin{\not=}{_A\|}(H\|_B)$.
Since I deal with guarded recursion only, $dex$ is well-defined.
\begin{table}[bt]
\caption{Operational Petri net semantics of CCSPS}
\label{tab:PN-CCSPS}
\vspace{-1ex}
\normalsize
\begin{center}
\framebox{$\begin{array}{@{}cc@{}}
\{a \triangleright \sum_{i \in I}a_i P_i\},\,\emptyset \ar{a_i} dex(P_i) &
\{\sum_{i \in I}a_i P_i\},\,\emptyset \ar{a_i} dex(P_i) \qquad (i\in I)\\[2ex]
\emptyset,\{a \triangleright \sum_{i \in I}a_i P_i\} \ar{a} \emptyset &
\displaystyle \frac{H,\,V \ar{a} J \qquad K,\,W \ar{a} L}
              {H\|_A \mathord\cup {_A\|}K ,\, V\|_A \mathord\cup {_A\|}W  \ar{a} J\|_A \mathord\cup {_A\|}L}~~
                                 (a\mathord\in A) \\[4ex]
\displaystyle \frac{H,\,V \ar{a} J}{H\|_A,\, V\|_A \ar{a} J\|_A}~~(a\mathop{\notin}A) &
\displaystyle \frac{H,\,V \ar{a} J}{_A\|H,\, {}_A\|V \ar{a} {_A\|}J}~~(a\mathop{\notin}A) \\[4ex]
\displaystyle\frac{H,\,V \ar{a} J}{\tau_I(H),\,\tau_I(V) \ar{\tau_I(a)} \tau_I(J)} &
\displaystyle\frac{H,\,V \ar{a} J}{f(H),\,f(V) \ar{f(a)} f(J)}
\end{array}$}
\end{center}
\vspace{-1ex}
\end{table}

Following \cite{Ol87}, I construct the unmarked Petri net $(S,T,F,R,\ell)$ of CCSPS
with $S:=\SC_{\rm CCSPS}$, specifying the tuple $(T,F,R,\ell)$ as a quaternary relation
$\mathord{\rightarrow} \subseteq \nat^S\times \nat^S\times \Act\times \nat^S$.
An element \plat{$H,V \ar{a} J$} of this relation denotes a transition
$t\mathbin\in T$ with $\ell(t)\mathbin=a$
such that $\precond{t}\mathbin=H$, $\widehat{t}\mathbin=V$ and $\postcond{t}\mathbin=J$.
The transitions \plat{$H,V,\ar{\alpha}J$} are derived from the rules of \tab{PN-CCSPS}.

Note that there is no rule for recursion. The transitions of an agent identifier $K$
are taken care of indirectly by the decomposition $dex(K) = dex(P)$,
which expands the decomposition of a recursive call into a decomposition of
an expression in which each agent identifier occurs within a guarded choice.

Trivially, the Petri net $\denote{P}$ associated to any CCSPS expression $P$ is finitary,
provided that all index sets $I$ for guarded choices occurring in $P$ are countable.

Olderog \cite{Ol91} shows that this operational semantics is consistent with the denotational
one of Defs.~\ref{df:parallel}, \ref{df:relabelling} and~\ref{df:guarded choice}, in the sense that
$\denote{P\|_A Q} \equiv \denote{P} \|_A \denote{Q}$---here the left-hand side follows the
operational semantics, and the right-hand side employs the denotational semantics of $\|_A$---and
similarly for the other operators. Moreover $\denote{K} \equiv \denote{P}$ for each agent identifier
with defining equation \plat{$K\stackrel{\it def}=P$}. Olderog's proof generalises smoothly to the
addition of read arcs. Here $\equiv$ is a non-transitive relation that Olderog calls \emph{strong bisimilarity}.

In \cite{vG15} I define \emph{structure preserving bisimilarity} on nets,
and show that it contains Olderog's strong bisimilarity. I also show that structure preserving
bisimilarity is congruence for the operators of CCSP that respects \emph{inevitability} when using
justness as completeness criterion. This means that if two systems are equivalent, and
in one the occurrence of a certain action is inevitable, then so is it in the other,
This implies that structure preserving bisimilarity is included in just must-testing equivalence $\equiv^J_{\rm must}$.
Hence $\denote{P\|_A Q} \equiv^J_{\rm must} \denote{P} \|_A \denote{Q}$, and similarly for the other operator,
i.e., the consistency of the operational and denotational semantics of Petri nets also holds up to
just must-testing equivalence.

\section{The just failures preorder is a congruence for $\|_A$}\label{congruence}

\begin{theoremRC}{\cite{Vog02}}{congruence parallel}
$\sqsubseteq^J$ is a precongruence for parallel composition.
\end{theoremRC}

\begin{proof}
Let $\sigma,\rho\in \Act^\infty$ and $A \subseteq \Act$. Then $\sigma\|_A \rho \subseteq \Act^\infty$
denotes the set of sequences of actions for which is it possible to mark each action occurrence 
as \emph{left}, \emph{right} or both, obeying the restriction that an occurrence of action $a$ is
marked both left and right iff $a \mathbin\in A$, such that the subsequence of all left-labelled action
occurrences is $\sigma$ and the subsequence of all right-labelled action occurrences is~$\rho$.

Obviously, $\nu \in \Act^\infty$ is the trace of an execution path $\pi$ of a net $N_1 \|_A N_2$
iff $\nu \in \sigma\|_A \rho$ for some traces $\sigma$ and $\rho$ of execution paths $\pi_1$ and
$\pi_2$ of $N_1$ and $N_2$, respectively.

Let $\pi = M_0 t_1 M_1 t_2 \dots$. Each transition $t_j$ has the form $(u_j,*)$, $(u_j,v_j)$ or $(*,v_j)$.
Moreover, each marking $M_i$ has the form \plat{$M_i^L \mathbin{\dcup} M_i^R$},
where $M_i^L$ is a marking of $N_1$ and $M_2^L$ is a marking of $N_2$.
If $t_{j+1}=(*,v_{j+1})$ then $M_j^L = M_{j+1}^L$.
Now $\pi_1$ can be obtained from $\pi$ by dropping all entries $(*,v_j)M_j$, and replacing the
remaining $t_j$ by $u_j$ and the remaining $M_i$ by $M_i^L$---I call it the \emph{projection} of
$\pi$ on $N_1$.
Likewise $\pi_2$ can be obtained from $\pi$ by dropping all entries $(u_j,*)M_j$, and replacing the
remaining $t_j$ by $v_j$ and the remaining $M_i$ by $M_i^R$.
\vspace{1ex}

\noindent
{\it Claim 1:} If $\pi[t\rangle$ and $t=(u,*)$ or $t=(u,v)$ then $\pi_1[u\rangle$.

\noindent
{\it Claim 2:} If $\pi[t\rangle$ and $t=(*,v)$ or $t=(u,v)$ then $\pi_2[v\rangle$.

\noindent
{\it Claim 3:} If $\pi_1[u\rangle$ and $\ell_1(u)\notin A$ then $\pi[(u,*)\rangle$.

\noindent
{\it Claim 4:} If $\pi_2[v\rangle$ and $\ell_2(v)\notin A$ then $\pi[(*,v)\rangle$.

\noindent
{\it Claim 5:} If $\pi_1[u\rangle$, $\pi_2[v\rangle$ and $\ell_1(u)=\ell_2(v)\in A$ then $\pi[(u,v)\rangle$.
\vspace{1ex}

\noindent
{\it Proof of Claim 1.} Suppose $\pi[t\rangle$ and $t=(u,*)$ or $t=(u,v)$.
Then $M_k[ t\rangle$ for some $k$ and 
$(\precond{t} + \widehat t\,) \cap \precond{t_{j+1}} = \emptyset$ for all $k \leq j < \length(\pi)$.
Now $M_k^L[u\rangle$. Moreover, for each $k \leq j < \length(\pi)$
    such that $t_j$ has the form $(u_j,*)$ or $(u_j,v_j)$ it follows that
    $(\precond{u} + \widehat u\,) \cap \precond{u_{j+1}} = \emptyset$.
    This implies that $\pi_1[u\rangle$.
\hfill \rule{5pt}{5pt}\vspace{1ex}%

\noindent
The proof of Claim 2 follows by symmetry. The remaining claims are obvious.
\vspace{1ex}

\noindent
    {\it Claim 6:} \(\F^J\!(N_1\|_A N_2) \mathbin= \left\{(\nu,\overline B) \left|
    \begin{array}{l@{}}
    \exists (\sigma,\overline C) \mathbin\in \F^J(N_1).~
    \exists (\rho,\overline D) \in \F^J(N_2).\\
    \nu \mathbin\in \sigma\|_A \rho \mathop\wedge C\mathop{\cap} D \mathop{\cap} A \mathbin\subseteq B \mathop\wedge (C \mathop\cup D){\setminus} A \mathbin\subseteq B
    \end{array}\right.\right\}\!.\)
\vspace{1ex}

\noindent
{\it Proof.} Let $(\nu,\overline B)\in \F^J(N_1\|_A N_2)$. Then $\nu=\trace(\pi)$ for an
execution path $\pi$ of $N_1\|_A N_2$, such that whenever $\pi[t\rangle$
then $\ell(t)\in B$. Define $\pi_1$ and $\pi_2$ as above, and let
$\sigma:=\trace(\pi_1)$, $\rho:=\trace(\pi_2)$,
$C := \{\ell_1(u) \mid \pi_1[u\rangle\}$
and $D := \{\ell_2(v) \mid \pi_2[v\rangle\}$.
By Defs.~\ref{df:path} and~\ref{df:failures}, $(\sigma,\overline{C})\in \F^J(N_1)$
and $(\rho,\overline{D})\in \F^J(N_2)$. Furthermore $\nu \in \sigma\|_A \rho$.

Now suppose $a \in C \cap D \cap A$. Then there are transitions $u$ and $v$ with
$\pi_1[u\rangle$, 
$\pi_2[v\rangle$
and $\ell_1(u)=\ell_2(v) = a \in A$.
By Claim 5, $\pi[(u,v)\rangle$. Thus $a = \ell((u,v)) \in B$.

Finally suppose $b \in (C \cup D){\setminus} A$. By symmetry I may restrict attention to the case
that $b \in C{\setminus}A$. Then there is a transition $u$ with
$\pi_1[u\rangle$ and $\ell_1(u)=b\notin A$.
By Claim 3, $\pi[(u,*)\rangle$. Thus $b = \ell((u,*)) \in B$.

Now let $(\sigma,\overline C) \mathbin\in \F^J(N_1)$,
$(\rho,\overline D) \mathbin\in \F^J(N_2)$,
$\nu \mathbin\in \sigma\|_A \rho$ and $B\mathbin\subseteq \Act$ satisfying
$C\cap D \cap A \mathbin\subseteq B$ and $(C \cup D){\setminus} A \subseteq B$.
Let $\pi_1$ and $\pi_2$ be execution paths of $N_1$ and $N_2$, respectively,
such that $\trace(\pi_1)=\sigma$,  $\trace(\pi_2)=\rho$, 
$\pi_1[u\rangle \Rightarrow \ell_1(u)\in C$ and
$\pi_2[v\rangle \Rightarrow \ell_2(v)\in D$.
Let  $\pi$ be an execution path of $N_1 \|_A N_2$ with $\trace(\pi)=\nu$,
such that its projections are $\pi_1$ and $\pi_2$.
Suppose $\pi[t\rangle$.
It remains to show that $\ell(t)\in B$.

First suppose $t$ has the form $(u,*)$. Then $\ell_1(u)\notin A$ and $\pi_1[u\rangle$ by Claim 1.
It follows that $\ell(t)=\ell_1(u) \in C{\setminus}A \subseteq B$.

The case that $t$ has the form $(*,v)$ proceeds likewise.

Finally suppose that $t=(u,v)$. Then $\ell(t)=\ell_1(u)=\ell_2(v)\in A$.
By Claims 1 and 2, one obtains $\pi_1[u\rangle$ and $\pi_2[v\rangle$.
Thus $\ell(t) \in C \cap D \cap A \subseteq B$. \hfill \rule{5pt}{5pt}\vspace{1ex}%

The theorem follows immediately from Claim 6 and \pr{failures}.
\qed
\end{proof}

\section{The just must-testing preorder contains the just failures preorder}\label{if}

\begin{propositionR}{just must testing}
  $N \sqsubseteq_{\rm must}^{\it J} N'$
  if
  $N \sqsubseteq^J N'$.
\end{propositionR}
 
\begin{proof}
  Suppose $N \sqsubseteq^J N'$ and let $\T$ be a test.
  Let $\pi'$ be an unsuccessful $\emptyset$-just execution path of $\T\|_\Act N'$.
  It suffices to find an unsuccessful $\emptyset$-just execution path of $\T\|_\Act N$.

Let $\pi_\T$ and $\pi_{N'}$ be the projections of $\pi$ to execution paths of $\T$ and $N$,
respectively, as defined in the proof of \thm{congruence parallel}. Let
$\sigma:=\trace(\pi_\T)$, $\rho:=\trace(\pi_{N'})$,
$C := \{\ell_\T(u) \mid \pi_\T[u\rangle\}$
and $D := \{\ell_{N'}(v) \mid \pi_{N'}[v\rangle\}$.
By Defs.~\ref{df:path} and~\ref{df:failures}, $(\sigma,\overline{C})\in \F^J(\T)$
and $(\rho,\overline{D})\in \F^J(N')$.
As in the first part of the proof of Claim 6 in the proof of \thm{congruence parallel},
but taking $B:=\emptyset$, one obtains $C \cap D \cap A = \emptyset$ and $(C \cup D){\setminus}A = \emptyset$.
Moreover, $\nu := \trace(\pi) \in \sigma \|_\Act \rho$.

  By \pr{failures}, $\F^J(N) \supseteq \F^J(N')$, so $(\rho,\overline{D})\in \F^J(N)$.
  Let $\pi_N$ be a execution path of $N$ such that $\trace(\pi_N)=\rho$ and
  $\pi_N[v\rangle \Rightarrow \ell_N(v)\in D$.
  Now compose $\pi_\T$ and $\pi_N$ into an execution path $\pi$ of $\T \|_\Act N$,
  such that $\trace(\pi)=\nu$ and its projections are $\pi_\T$ and $\pi_N$.
As in the second part of the proof of Claim 6 in the proof of \thm{congruence parallel},
but taking $B:=\emptyset$, one obtains $\pi[t\rangle \Rightarrow \ell(t)\in \emptyset$.
It follows that $\pi$ is $\emptyset$-just.
Moreover, as $\pi'$ is unsuccessful, so is $\pi_\T$, and hence also $\pi$.
\qed
\end{proof}

\section{Qualitatively timed must-testing}
\label{timed}

Let $N$ be a finitely branching safe net.
For each execution path $\pi$ of $N$, I define the \emph{slowest} timed execution path $\widetilde\pi$
through the following algorithm, which uses the variable $\hat\pi$ to
store the suffix of $\pi$ that still needs to be executed.
$\widetilde \pi$ starts out empty, and $\hat\pi:=\pi$.\vspace{-1ex}
\begin{enumerate}[(1)]
\item Let 1 unit of time pass, i.e., add a time step $1$ to $\tilde\pi$.
  Now finitely many transitions are enabled in the current marking. Store those in the set $T_{\it en}$.
\item As long as $T_{\it en} \neq \emptyset$, fire the first transition of $\hat \pi$, i.e., add it
  to $\tilde\pi$; and remove this
  first transition from $\hat \pi$; remove from $T_{\it en}$ all transitions that are no longer enabled.
\item When $T_{\it en} = \emptyset$, go to (1).
\end{enumerate}
The constructed path $\widetilde \pi$ arises in the limit.
\begin{lemmaR}{infty}
If $\pi$ is just, then $\zeta(\widetilde\pi)=\infty$.
\end{lemmaR}
\begin{proof}
  In case $\pi$ is infinite, this is obvious, since the algorithm keeps adding time-1 steps regularly.
  In case $\pi$ in finite, in its final marking no further transitions are enabled.
  The algorithm will now continue to add time-1 steps forever.
  \qed
\end{proof}
Given a finite execution path $\pi$, let $\widetilde\pi^*$ be obtained from $\widetilde\pi$ by
leaving out all trailing time-steps. When calculating $\widetilde\pi^*$, the algorithm simple stops
as soon as the last transition of $\hat\pi$ has been fired.
\begin{lemmaR}{finite}
Let $|\pi|$ denotes the number of transitions in $\pi$. Then $\zeta(\widetilde\pi^*) \leq |\pi|$.
\end{lemmaR}
\begin{proof}
   This follows because at least one transition must be scheduled between each two time steps.
  \qed
\end{proof}

\begin{lemmaR}{slower}
Each timed execution path $\chi$, ending with a transition, can be transformed in an untimed execution path $\theta$,
namely by omitting the lapses of time that are recorded in $\chi$. Now $\zeta(\widetilde\theta^*) \geq \zeta(\chi)$.
\end{lemmaR}
\begin{proof}
This follows because when sticking to the order of transitions in $\chi$, the timed execution path
$\widetilde\theta^*$ schedules each transition as late as possible.
\qed
\end{proof}

\begin{lemmaR}{timed}
  Let $N$ be a finitely branching safe net.
  Then each just execution path of $N$ contains a transition labelled $w$
  iff there is a duration $D \in\reals_0^+$ such that a transition labelled $w$ occurs in each
  timed execution path $\chi$ of $N$ with $\zeta(\chi)>D$.
\end{lemmaR}

\begin{proof}
Suppose there is a duration $D \in\reals_0^+$ such that a $w$-transition occurs in each
timed execution path $\chi$ of $N$ with $\zeta(\chi)>D$. Let $\pi$ be a just execution path of $N$.
Then $\zeta(\widetilde\pi)\mathbin=\infty$ by \lem{infty}, so $w$ occurs in $\widetilde\pi$, and hence in $\pi$.

Now suppose each just execution path of $N$ contains a $w$-transition.
For each just execution path $\pi$ of $N$, let $\overline\pi$ the prefix of
$\pi$ up to and including the first occurrence of a $w$-transition, and let 
$|\overline\pi|$ be the number of transitions in $\overline\pi$. By K\"onig's Lemma,
using that $N$ is finitely branching, there exists a finite upper bound $D$ on all the values $|\overline\pi|$.
Now $\zeta(\widetilde{\overline{\pi}}^*)\leq |\overline\pi| \leq D$ by \lem{finite}.

Let $\chi'$ be a timed execution path of $N$ with $\zeta(\chi')>D$.
Then there is a finite prefix $\chi$ of $\chi'$ with $\zeta(\chi)>D$.
By \thm{feasible} its untimed version $\theta$ must
be a prefix of a just execution path $\pi$, and since \plat{$\zeta(\widetilde{\theta}^*) \geq \zeta(\chi) > D$} by
\lem{slower}, it follows that \plat{$\widetilde{\overline\pi}^*$} must be a prefix of \plat{$\widetilde{\theta}^*$}. Since
a $w$-transition occurs in $\widetilde{\pi}^*$,  it also occurs in \plat{$\widetilde\theta^*$}, and hence in $\chi$.
\qed
\end{proof}

\noindent
\textit{Proof of \thm{must eventually}}.
Let $N,N'$ be finitely branching safe nets.

``Only if'': Suppose $N \not\sqsubseteq^J N'$.
By \pr{failures} there is an $(\sigma,X) \in \F(N'){\setminus}\F(N)$.
Let $\T$ be the universal test for the just failure pair $(\sigma,X)$, displayed in \fig{universal test}.
Then $N$ \textbf{must} $\T$ but $\neg(N'$ \textbf{must} $\T).$
Hence each just execution path of $\T\|_\Act N$ is successful, but there is a just execution path of
$\T\|_\Act N'$ that is unsuccessful.
Since $N,N'$ are finitely branching safe nets, so are $\T\|_\Act N$ and $\T\|_\Act N'$.

Each just execution path $\pi$ of $\T\|_\Act N$ will reach a marking where a transition $t$ labelled $w$ is enabled.
Given the shape of $\T$, no other transition of $\T\|_\Act N$ can remove the token from the unique
preplace of $t$. Hence $\pi$ must contain transition $t$.
Thus \lem{timed} implies that $N$ must eventually pass $\T$.
On the other hand, $\T\|_\Act N'$ has a just execution path that does not contain a transition
labelled $w$, so \lem{timed} denies that $N'$ must eventually pass $\T$.
It follows that $N \not\sqsubseteq_{\rm must}^{\rm ev.} N'$.

``If'': Suppose $N \not\sqsubseteq_{\rm must}^{\rm ev.} N'$.
Then there is a finitely branching safe test $\T'$, such that $N$ must eventually pass $\T'$, yet
$N'$ does not. Modify $\T'$ into $\T$ by replacing each $w$-transition by a sequence of a $\tau$- and
a $w$-transition, with a single place in between.
If each timed execution path of $\T'\|_\Act N$ will reach a $w$-transition within time $D$, then 
each timed execution path of $\T\|_\Act N$ will reach a $w$-transition within time $D{+}1$.
Hence $N$ must eventually pass $\T$, yet $N'$ does not. 

Since $N,N'$ and $\T$ are finitely branching safe nets, so are $\T\|_\Act N$ and $\T\|_\Act N'$.
By \lem{timed}, each just execution path of $\T\|_\Act N$ contains a $w$-transition,
and thus $N$ \textbf{must} $\T$. On the other hand, $\T\|_\Act N$ has a just execution path $\pi$
that does not contain a $w$-transition. Given the shape of $\T$, path $\pi$ does not contain a
marking where a $w$-transition is enabled, for it it did, justness would force that transition to
occur in $\pi$. It follows that $N \not\sqsubseteq_{\rm must}^J N'$, and hence $N \not\sqsubseteq^J N'$.
\qed

\end{document}